\documentclass[a4paper,11pt]{article}
\pdfoutput=1
\usepackage{jheppub}
\usepackage{amsmath}
\usepackage{amssymb}
\usepackage{color}
\usepackage{graphicx}
\usepackage{hyperref}
\usepackage{xspace,slashed}
\usepackage[utf8]{inputenc}
\usepackage{subcaption}
\usepackage{multirow}
\usepackage{algorithm}
\usepackage{algorithmic}
\usepackage{stackengine}

\usepackage{soul}

\usepackage{booktabs}

\makeatletter
\g@addto@macro\bfseries{\boldmath}
\makeatother

\newcommand{\order}[1]{\mathcal{O}\left(#1\right)}

\newcommand{\fb}{\;\mathrm{fb}}

\newcommand{\GeV}{\;\mathrm{GeV}}
\newcommand{\TeV}{\;\mathrm{TeV}}
\newcommand{\as}{\alpha_s}

\newcommand{\nott}{{\!\not\, t}}
\newcommand{\pTt}{p_{T,t}}
\newcommand{\mtop}{m_\text{top}}
\usepackage{xcolor}
\definecolor{light-gray}{gray}{0.8}

\definecolor{semiblue}{rgb}{0.3,0.3,0.8}
\newcommand{\logbook}[2]{}

\newcommand{\ttbar}{{t\bar t}}
\newcommand{\tbar}{{\bar t}}
\newcommand{\dytt}{\Delta y_{\ttbar}}
\newcommand{\HT}{H_T^{\ttbar+\text{jets}}}

\newcommand{\OXaff}{Rudolf Peierls Centre for Theoretical Physics,
  Clarendon Laboratory, Parks Road, Oxford OX1 3PU, UK}
\newcommand{\ASCaff}{All Souls College, Oxford OX1 4AL, UK}
\newcommand{\WADaff}{Wadham College, Oxford OX1 3PN, UK}

\preprint{
  \begin{flushright}
    OUTP-21-01P
  \end{flushright}
}

\title{
  Framing energetic top-quark pair production at the LHC
}
\author[a,b]{Fabrizio Caola,}
\author[a]{Fr\'ed\'eric A. Dreyer,}
\author[a]{Ross W.\ McDonald,}
\author[a,c]{Gavin P.\ Salam}
\affiliation[a]{\OXaff}
\affiliation[b]{\WADaff}
\affiliation[c]{\ASCaff}

\abstract{
  Top-quark pair production is central to many facets of LHC
  physics. At leading order, the top and anti-top are produced in a
  back-to-back topology, however this topology accounts only for a
  minority of $t \bar t$ events with TeV-scale momentum transfer.
  The remaining events instead involve the splitting of an initial
  or final-state gluon to $t \bar t$. We provide simple quantitative
  arguments that explain why this is the case,
  and examine the interplay between different topologies and a range
  of variables that characterise the event hardness.
  We then develop a method to classify the topologies of individual
  events and use it to illustrate our findings in the context of
  simulated events, using both top partons and suitably defined
  fiducial tops.
  For events with large $t \bar t$ invariant mass, we
  comment on additional features that have important experimental and
  theoretical implications.  }

\begin{document}

\maketitle 
\section{Introduction}

Top quarks are among the most central objects to collider physics
today.
As the heaviest known fundamental particle, the top-quark has a unique
place in the standard model (SM), the only particle with a Yukawa
coupling close to unity, making it a key player in many Beyond
Standard Model (BSM) scenarios and in discussions of the stability of
the SM universe~\cite{Degrassi:2012ry}.
Top-quark pair production is an increasingly important input to fits of
parton distribution functions (PDFs)~\cite{Alekhin:2017kpj,Ball:2017nwa,Hou:2019qau,Bailey:2020ooq,Amat:2019upj,Bailey:2019yze,Czakon:2019yrx}.
It plays a crucial role in effective field theory (EFT)
studies~\cite{AguilarSaavedra:2018nen,Hartland:2019bjb,Brivio:2019ius,Bissmann:2020mfi,Ellis:2020unq},
provides an avenue to learning about the top-quark Higgs Yukawa
coupling~\cite{Sirunyan:2020eds}.
More generally top quarks are
omnipresent in beyond standard-model (BSM) searches,
both as signal objects and for backgrounds (e.g.\ as a source of
leptons and $b$ quarks).

Early studies of top-quark production at Fermilab's Tevatron and CERN's
Large Hadron Collider (LHC) were
restricted to configurations where the top quarks had a transverse
momentum that was comparable to the top-quark mass.
Today, at each of the ATLAS and CMS collision points, the LHC has
produced over a hundred thousand events with top quarks having a
transverse momentum $p_T > 500\GeV$ and a couple of thousand events with
$p_T > 1\TeV$ (all decay channels combined), with those numbers
expected to increase by a factor of $25{-}30$ at the high-luminosity
LHC~\cite{Azzi:2019yne}.%
\logbook{2390240}{cf. Fig.68, top-right, based on NNLO calculation;
  13/14TeV cross sections with $p_T>1\TeV$, summed over all decays,
  are $19.9fb$ and $27.6fb$ according to Pythia
\begin{verbatim}
./cross-section -pythia8 -out a13 -ttbar -ptmin_gen 1000 -noUE -parton -noISR -noFSR -nev 1e4 -rts 13000
\end{verbatim}
  and they are $1240\fb$ and $1580\fb$ for $p_T>500\GeV$, i.e.\ about
  60 times larger.
  Looking at the average top $p_t$ at NLO, we have
\begin{verbatim}
  mergeidx.pl err_hist:truth:all:ptavg:1000-inft:delta_y -f 13TeV_powheg_parton_rapmax99_allrseq.dat
\end{verbatim}
  and see that we get a cross section of 2.628184e-06nb=2.63fb 
  including the branching fraction to muons on one side ($4/27$),
  which would then give a total cross section of $17.8\fb$.
  That gives us about $2,500$ events for $140\fb^{-1}$ and $53,000$
  for $3000\fb^{-1}$.
  Including a branching fraction of $4/27$ for the semi-muonic
  channel, those reduce to about $370$ and $8000$ respectively.
  Multiplying those numbers by $60$ to get $p_T>500\GeV$ gives: $150k$
  total $\ttbar$ events and $22k$ in the semi-muonic channel.
}
These large numbers of events provide an opportunity for percent-level
precision across a wide kinematic range, with corresponding
experimental measurements well underway~%
\cite{%
  Aaboud:2018eqg,
  Aad:2019ntk,
  Aad:2019hzw,
  Sirunyan:2020vwa,
  Sirunyan:2019zvx,
  Sirunyan:2018wem
}.
This measurement programme will benefit a wide range of
physics areas.

To reap these benefits, it will be necessary to achieve corresponding
percent-level understanding of theoretical predictions and one may
expect to leverage the impressive ongoing progress in perturbative
calculations (for a review see Ref.~\cite{Heinrich:2020ybq}).
The reliability of this approach depends, however, on an assumption
that the perturbative series is well behaved.
A prerequisite is that the event topology that appears at LO, i.e.\ a
$2\to2$ process with a back-to-back top-pair, dominates over topologies that start to arise
only at higher orders, for example a boosted $t\tbar$ system recoiling against
a hard jet.
As we shall see, the extent to which this is true depends on the
choice of observable used to characterise the event hardness.
For some choices, the hierarchy holds as expected, but for other
widely used choices of event hardness scale this is not the case,
(e.g.\ for any observable that sums the transverse momenta of both jets and the
top quarks).
Accordingly, it is crucial to develop an understanding of the interplay
between various widely used measures of the event hardness and the
underlying topology of $t\bar t$ production.
This is one of the main goals of this manuscript.

There are multiple potential benefits to having this understanding.
For applications where precision is crucial, it can inform the choice
of measurement observables.
More generally, we will develop an approach for identifying the
topology of any given event, so as to be able to separate out
subsets of events that may probe different underlying physics in
the specific application being considered, whether a PDF fit, an EFT
fit or a BSM search.

This article is structured as follows.
In Section~\ref{sec:theory}, we examine a range of variables used to
characterise the hardness of $\ttbar$ events, examine the event
topologies that can be produced at leading order (LO) and
next-to-leading order (NLO), give a simple analysis of our
expectations for their relative sizes, and finally consider the
interplay between these topologies and the event hardness variables.
In Section~\ref{sec:parton-level-analysis} we introduce a procedure
for identifying the topologies of individual events given the
top quarks and a list of the other event particles, apply the
procedure to simulated $\ttbar$ events, and compare the results to the
expectations of Section~\ref{sec:theory}.
In Section~\ref{sec:hadronic-analysis} we combine this analysis with a
fiducial reconstructed top-quark definition designed to identify
top-quark candidates from the full set of final-state particles across
a wide range of transverse momenta. 
In Section~\ref{sec:conclusions}, we conclude with a discussion of the
implications of our results.
Throughout this paper we will consider the semi-leptonic $\ttbar \to
b\bar b \mu \nu_\mu jj$ decay channel, though the arguments that we
make apply generally to any $\ttbar$ decay channel.

\section{Theoretical considerations}
\label{sec:theory}

In this section we review various event-hardness measures, and discuss
some basic expectations about their behaviour for events that involve
large momentum-transfer and contain a $\ttbar$ pair.

\subsection{Event hardness variables and their leading-order
  behaviour}
\label{sec:hardness-scales}

\begin{table}
  \centering
  \begin{tabular}{ccl}
    \toprule
    Hardness variable && explanation\\
    \midrule
    $p_{T}^{\text{top,had}}$
          && transverse momentum of hadronic top candidate
    \\
    $p_{T}^{\text{top,lep}}$
          && transverse momentum of leptonic top candidate
    \\
    $p_{T}^{\text{top,max}}$
          && $p_T$ of the top (anti-)quark with larger $m_T^2 = p_T^2+m^2$
    \\
    $p_{T}^{\text{top,min}}$
          && $p_T$ of the top (anti-)quark with smaller $m_T^2 = p_T^2+m^2$
    \\
    $p_{T}^{\text{top,avg}}$
          && $ \frac12(p_{T}^{\text{top,had}} + p_{T}^{\text{top,lep}}) $
    \\
    \midrule
    $\frac12 H_T^{\ttbar}$  
          && with $H_T^{\ttbar} = m_T^\text{top,had} + m_T^\text{top,lep}$
    \\
    $\frac12 \HT$ 
          && with $\HT = m_T^\text{top,had} + m_T^\text{top,lep} + \sum_i p_T^{j_{\nott,i}}$
    \\
    $m_T^{J,\text{avg}}$
          && average $m_T$ of the two highest $m_T$ large-$R$ jets ($J_1,J_2$)
    \\
    \midrule
    $\frac12 m^{\ttbar}$
          && half invariant mass of $p^\ttbar = p^\text{top,had} + p^\text{top,lep}$
    \\
    \midrule
    $p_{T}^{\ttbar}$
          && transverse component of $p^\ttbar$
    \\
    $p_T^{j_{\nott,1}}$
          && transverse momentum of the leading small-$R$ non-top jet
    \\
    \bottomrule
  \end{tabular}
  \caption{Variables that may be used to characterise a hard kinematic
    scale
    in events with a semi-leptonically decaying $\ttbar$ pair.
    All observables within a given group are identical to each other at
    leading order.
    The $j_{\nott,i}$ jets correspond to $R=0.4$ non-top jets, while
    the $J_i$ jets
    corresponds to large-$R$ jets (whose clustering inputs include the
    top quarks).
    Further details about the jet finding are given in
    Sections~\ref{sec:parton-level-analysis} and \ref{sec:hadronic-analysis}.
  }
  \label{tab:hard-scale-choices}
\end{table}

We start by examining measures to characterise large
momentum transfer in $\ttbar$ events, i.e.\ the event
hardness, including a discussion of their leading-order
distributions. 
A wide variety of such measures is used in the literature and we
provide an illustrative selection of them in
Table~\ref{tab:hard-scale-choices}, organised into groups that are
identical at LO, i.e.\ for events that consist of just a single
back-to-back $\ttbar$ pair.

The first set of observables simply measures the top-quark transverse
momentum.
They differ only in terms of which top-quark is used, which is why
they are identical at LO (order $\as^2$).
We also have an all-order relation between some of the observables,
specifically\footnote{The rightmost
  expression of our 
  Eq.~(\ref{eq:pttop-equivalences}) is referred to as
  $d\sigma/dp_{T,\text{avt}}$ in Ref.~\cite{Czakon:2016dgf}.
  This differs from our $d\sigma/dp_T^\text{top,avg}$ beyond LO.}
\begin{equation}
  \label{eq:pttop-equivalences}
  \frac{d\sigma}{dp_{T}^{\text{top,had}}}
  =
  \frac{d\sigma}{dp_{T}^{\text{top,lep}}}
  =
  \frac12 \left(
    \frac{d\sigma}{dp_{T}^{\text{top,max}}}
    +
    \frac{d\sigma}{dp_{T}^{\text{top,min}}}
  \right)\,.
\end{equation}
Note that we have chosen to define the ``max/min'' based on the value
of $m_T$ rather than $p_T$, but results would be essentially unchanged
if we instead used $p_T$.
In the LO calculations that we report below, it will be convenient to have
the shorthand $\pTt$ for the transverse momentum of either of the
top quarks.

The next set of observables provides measures of the hard scale of the
event that at LO include the top-quark mass and transverse momentum.
The $H_T^\ttbar$ variable is identical to the $H_T$ of Czakon, Heymes
and Mitov~\cite{Czakon:2016dgf} and of Catani et al.~\cite{Catani:2019iny}.
The $\HT$ observable provides a democratic evaluation of the event
hardness across all objects.
At high scales is very similar to the
$m_\text{eff}$ variable used in supersymmetry searches, which
is the scalar sum of the transverse momenta of all jets, leptons and
missing momenta, see e.g.\ Ref.~\cite{Aad:2021zyy}.\footnote{The
  reason for not directly including $m_\text{eff}$ in this paper is
  that it cannot be meaningfully used with top partons, but only
  with top decay products. }
The $m_T^{J,\text{avg}}$ quantity is based on large-$R$ jets, with the
details of the jet finding discussed in more detail below. 
It is not part of the standard set of $\ttbar$ event-hardness
measures, but we include it here because it gives a faithful
reflection of the hardness of the main $2\to 2$ scattering in the
event, regardless of the precise role played by the top quarks in that
scattering.

At leading order, in the limit where $p_{T,t}$ grows much larger than
the top-quark mass ($m_\text{top}$), all three observables in this group
become identical to the $\pTt$ type observables of the first group.
Structurally, for $\pTt \gg \mtop$ but still significantly smaller
than the collider centre-of-mass energy $\sqrt{s}$, the LO cross
section is given by
\begin{equation}
  \label{eq:sigma-pTt-LO}
  \frac{d\sigma}{d\pTt^2}
  = \frac{\as^2 \pi}{4\pTt^4} \left[
    c_{gg} \mathcal{L}_{gg}(4\pTt^2/s)
    +
    c_{q\bar q}
    \mathcal{L}_{q\bar q}(4\pTt^2/s)
  \right]
\end{equation}
Here the $c_{gg}$ and $c_{q\bar q}$ are numerical constants, of the
order of $0.1$, which depend on how steeply the PDFs fall with
increasing $x$.
They are discussed in Appendix~\ref{sec:LO-notes-pt}, together with
our specific definition the partonic luminosities $\mathcal{L}_{gg}$
and $\mathcal{L}_{q\bar q}$.
Since this is a LO calculation, for simplicity we have left out
explicit renormalisation and factorisation scale dependence.

The next observable in Table~\ref{tab:hard-scale-choices} is the invariant
mass of the $t\bar t$ system.
For $m_\ttbar \gg \mtop$, the LO distribution is
\begin{equation}
  \label{eq:sigma-mtt-LO}
  \frac{d\sigma}{dm_\ttbar^2} = \frac{\alpha_s^2 \pi}{m_\ttbar^4}
  \left[
    \left(\frac13 \ln \frac{m_\ttbar^2}{\mtop^2} - \frac7{12} \right)
   \mathcal{L}_{gg}(m_\ttbar^2/s)
    +
    \frac{8}{27}\, \mathcal{L}_{q\bar q}(m_\ttbar^2/s)
  \right].
\end{equation}
Relative to the result for the $\pTt$ distribution,
Eq.~(\ref{eq:sigma-pTt-LO}), a key structural difference here is the
presence of a factor $\ln \frac{m_\ttbar^2}{\mtop^2}$ multiplying the
gluon--gluon luminosity.
One can understand the origin of this logarithm by considering fixed
$m_\ttbar$ and examining the distribution of $\dytt$, the difference
in rapidity between the top and anti-top.
At large $\dytt$, the gluon-fusion contribution is dominated by a
$t$-channel top-quark exchange diagram and the cross section becomes a
constant, independent of $\dytt$. 
Integrating over $\dytt$ up to its kinematic limit,
\begin{equation}
  \label{eq:dytt-max}
  \dytt^{\max} = \cosh^{-1}\left(\frac{m_\ttbar^2}{2\mtop^2}-1\right)
  = \ln \frac{m_\ttbar^2}{\mtop^2} + \order{\frac{\mtop^2}{m_\ttbar^2}}\,,
\end{equation}
then yields the logarithmic factor seen in
Eq.~(\ref{eq:sigma-mtt-LO}), cf.\ Appendix~\ref{sec:LO-notes-mtt}.
The large-$\dytt$ enhancement of the gluon-fusion versus $q\bar q$
contributions provides a potentially powerful handle for separately
constraining the $q\bar q$ and $gg$ luminosities in PDF fits.

Note that at the large $\dytt$ values that dominate the gluon-fusion
contribution to Eq.~(\ref{eq:sigma-mtt-LO}), the top-quark transverse
momentum $\pTt$ is much smaller than its $\dytt=0$ value of
approximately $m_\ttbar/2$.
This is the reason why one should be wary of using $m_\ttbar$ as a
renormalisation and factorisation scale for calculating the $m_\ttbar$
distribution, and would expect the use of $\mu=m_\ttbar$ to lead to
poor stability, as observed in
Ref.~\cite{Czakon:2016dgf}.\footnote{Similar considerations apply to
  the measurement of the running top-quark mass as a function of
  $m_\ttbar$~\cite{Sirunyan:2019jyn}. Indeed, at high $m_\ttbar$ the
  sensitivity of the cross section to the top-quark mass will come
  predominantly from the region close to the kinematic limit of $\dytt$,
  where the largest virtuality of the produced or exchanged top quarks
  is much closer to $\mtop$ than to
  $m_\ttbar$.\label{ft:CMS-running-mass}}
In practice, once one uses a dynamical scale, the LO result is no
longer flat in $\dytt$, but is sensitive to the varying scale of both
$\as$ and the PDF, and typically results in a $\dytt$ distribution
that is peaked at large $\dytt$.\footnote{This feature was observed
  numerically in the context of FCC-hh studies in section 12.3 of
  Ref.~\cite{Mangano:2016jyj}.}
Some plots illustrating these points are given at LO in
Appendix~\ref{sec:LO-notes} and at NLO in
Section~\ref{sec:parton-level-results} and Appendix~\ref{sec:mttbar-beyond-LO}.

The importance of large $\dytt$ values makes the $m_\ttbar$
observable subtle both theoretically and experimentally.
The theoretical subtleties are discussed briefly in
Appendix~\ref{sec:mttbar-beyond-LO} and have two main facets: firstly,
for $t$-channel top-quark exchange a rich structure of logarithmically
enhanced terms, $\as^n \ln^m m_\ttbar/\mtop$, develops beyond LO.
Secondly, contributions from 4-top final states with $t$-channel gluon
exchange, and EW $b\bar b \to t\bar t$ diagrams with $t$-channel $W$
exchange both scale as $1/(\mtop^2 m_\ttbar^2)$ rather than
$1/(m_\ttbar^4)$.
For sufficiently large $m_\ttbar$ values they dominate over other
contributions.
At the LHC, they bring
only a small contribution, because of suppression by phase-space and
couplings, but this would no longer be the case at a $100\TeV$ $pp$
collider.

Experimentally there are also at least two facets to the subtleties
for the $m_\ttbar$ observable.
Firstly, large $\dytt$ implies large rapidities for at least one of
the two tops, which may then be beyond the detector acceptance,
notably for $b$-tagging and lepton identification.
Secondly, the spread of $\pTt$ values across the range of $\dytt$ adds
the complication that standard measurement approaches, which use either resolved
or boosted top reconstruction techniques, cannot reconstruct tops
across the whole range of $\dytt$.
We will return to this issue below.

The last two observables that we show in
Table~\ref{tab:hard-scale-choices}, $p_{T}^{\ttbar}$ and
$p_T^{j_{\nott,1}}$, are identically zero at order $\as^2$.
In the absence of $p_T$ and rapidity acceptance cuts for jets, the two
observables are identical at order $\as^3$, as long as one has a
concrete scheme that separates the top decay products from other event
particles.

\subsection{Topologies beyond LO}
\label{sec:topologies-beyond-LO}

Having reviewed the key characteristics of the LO distributions, we
can now turn to the main topic of this paper, the question of
topologies beyond LO and the interplay between topologies and event
hardness scales.

\begin{figure}
  \centering
  \includegraphics[width=\textwidth]{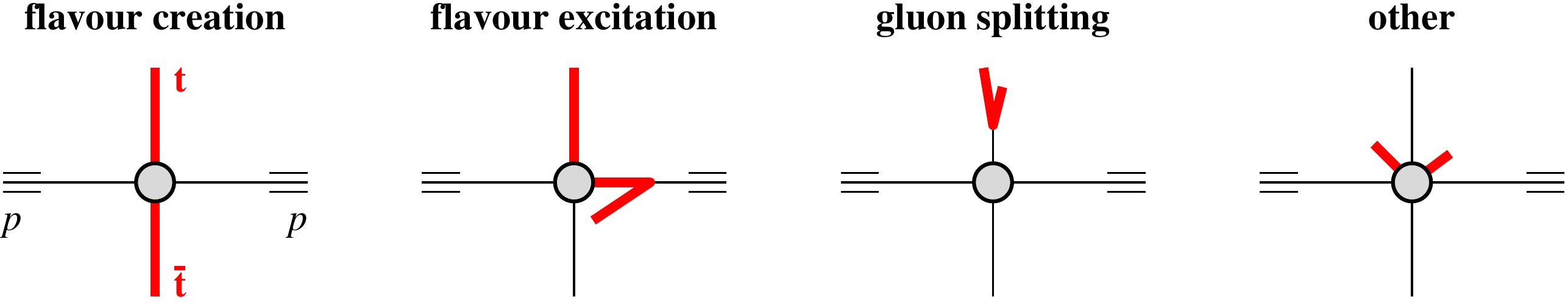}
  \caption{Illustration of classes of event topology for top
    production.
    Thick red lines represent top (anti-)quarks, while thin black
    lines represent light partons (quarks or gluons).
    Protons are depicted as entering from the left and right-hand
    sides.  }
  \label{fig:classes}
\end{figure}

The topologies that we will consider are illustrated in
Fig.~\ref{fig:classes} and will be familiar to some readers from older
discussions of $b$-quark production.
The flavour creation (FCR) configuration is the dominant mechanism for
top production at low transverse momentum.
It is the only topology that is present at leading order (LO) in a
strong-coupling perturbative expansion.
In flavour excitation (FEX), a $\ttbar$ pair can be produced by an
initial state splitting, with one of the pair undergoing a large
momentum-transfer scattering with a light parton.
Gluon splitting (GSP) involves production of a $t\bar t$ pair during
jet fragmentation.
Both FEX and GSP start at next-to-leading order (NLO).
Finally some events do not readily fall into any of these categories,
for example two high-transverse momentum light-flavour jets plus a
(relatively) soft additional gluon that splits to $\ttbar$.
These arise only at NNLO and beyond.

Relative to LO, the FEX and GSP topologies involve a
factor $\as \ln p_T/\mtop$, where $p_T$ is generally the transverse
momentum of the hardest object in the event.
The $\ln p_T/\mtop$ factor that arises at the LHC is typically not
large: e.g.\ for $p_T \sim 1\TeV$, it is of the order $2$, which would
not be expected to compensate for the extra power of $\as$ relative to
LO, and one might expect FEX and GSP to be small compared to FCR.%
\footnote{At a $100\TeV$ $pp$ collider, the logarithms can be larger,
  which might then at first sight explain the observation in section
  12.3 of Ref.~\cite{Mangano:2016jyj} that GSP contributes
  significantly to high-$p_T$ top production.}
As we shall see, this intuition misses important considerations.
To help understand this, Table~\ref{tab:channels} shows the different
factors that come into the calculation of the cross section for the
FCR, FEX and GSP topologies.
We consider a $2\to2$ hard scattering energy of $2\TeV$ and take the
case of $90$~degree scattering in the centre of mass, which dominates
high-$p_T$ production.
This corresponds to each outgoing object from the $2\to2$ scattering
having a transverse momentum of $1\TeV$ and identical rapidity.

\begin{table}
  \centering
  \begin{tabular}{cccccc}
    \toprule
    topology & channel & $|\text{ME}|^2$ & luminosity & FS splitting & product\\
    \midrule
    \multirow{2}{*}{FCR}
    & $gg \to \ttbar$       & 0.15 & 0.16   & 1 & 0.024 \\
    & $q_i\bar q_i \to \ttbar$  & 0.22 & 0.13   & 1 & 0.028 \\
    \midrule
    \multirow{2}{*}{FEX}
    & $tg \to tg$           & 6.11 & 0.0039 & 1 & 0.024 \\
    & $t\Sigma \to t\Sigma$ & 2.22 & 0.0170 & 1 & 0.038 \\
    \midrule
    & $gg \to gg(\to t\bar t)$           & 30.4 & 0.16 & $\mathcal{P}_{g \to \ttbar} \simeq 0.004$&0.020\\
    GSP
    & $g\Sigma \to g(\to t\bar t)\Sigma$ & 6.11 & 1.22 & $\mathcal{P}_{g \to \ttbar} \simeq 0.004$&0.031\\
    & $q\bar q \to gg(\to t\bar t)$      & 1.04 & 0.13 & $\mathcal{P}_{g \to \ttbar} \simeq 0.004$&0.001\\
    \bottomrule
  \end{tabular}
  \caption{Factors contributing to the top-production cross section
    for a variety of partonic scattering channels.  %
    In each case the $2\to 2$ squared matrix element ($|\text{ME}|^2$,
    with a $g^4 = (4\pi\as)^2$ factor 
    stripped off as in Eqs.~(\ref{eq:qqbar2qqbar}),
    (\ref{eq:qqp2qqp})) is given in the massless limit (valid when
    $p_T \gg m_t$), for $90^\circ$ scattering in the partonic
    centre-of-mass frame.
    The partonic luminosities, defined as in Eq.~(\ref{eq:lumi-def}),
    are given for a proton--proton centre of mass energy of
    $\sqrt{s} = 13\TeV$ and for producing a partonic system mass of
    $\sqrt{\hat s} = 2\TeV$.
    We set the factorisation scale to $\mu = 1\TeV$.
    $\Sigma$
    denotes a sum over all (non-top) quark and anti-quark flavours.
    The luminosities have been evaluated with the
    \texttt{PDF4LHC15\_nnlo\_mc}~\cite{Butterworth:2015oua} set,
    re-evolved in a
    six-flavour scheme with \texttt{HOPPET}~\cite{Salam:2008qg} using
    NNLO splitting and threshold-matching
    functions~\cite{Moch:2004pa,Vogt:2004mw,Buza:1995ie,Buza:1996wv}. 
    The final-state splitting probability $P_{g\to \ttbar}$ is
    obtained using Eq.~(\ref{eq:g2ttbar-semi-asymptotic}).
    The results in the final column are to be taken as order of
    magnitude estimates, illustrating the commensurate sizes of
    different channels.
    \logbook{30ffa45}{Raw results for luminosities to be found in
      ../asymptotic-limit/calculate-lumis.result.txt}
    \logbook{}{Run
      ../asymptotic-limit/g2QQbar-python-checks/wbb-checks.gp to get
      $P_{g\to \ttbar}\simeq 0.0041$, specifically via
      ``Coefficient of alpha_s TR/2pi for ptR=1000, m=173: 0.581120787673439''
    }
  }
  \label{tab:channels}
\end{table}

The first point that we highlight is that the underlying $2\to2$
matrix elements for the FCR process are an order of magnitude smaller
than for FEX and GSP.
To illustrate  the origin of this analytically in one simple case,
consider $90^\circ$ scattering 
in the limit $p_T \gg
\mtop$, and compare for example the squared matrix element
relevant for the $q_i\bar q_i \to t\bar t$ channel of FCR (cf.\ \cite{Combridge:1977dm}
or \cite{Ellis:1991qj}),
\begin{equation}
  \label{eq:qqbar2qqbar}
  \frac{1}{g^4}\sum_\text{spin,colour}
  |{\cal M}_{q\bar q \to q'\bar q'}|^2
  = \frac{C_F}{N_C}
  \frac{\hat t^2 + \hat u^2}{\hat s^2}
  = \frac{C_F}{N_C} \cdot \frac12\,,
\end{equation}
to that involved in the $q t \to q t$ channel of FEX,
\begin{equation}
  \label{eq:qqp2qqp}
  \frac{1}{g^4}\sum_\text{spin,colour}
  |{\cal M}_{q q' \to q q'}|^2
  = \frac{C_F}{N_C}
  \frac{\hat s^2 + \hat u^2}{\hat t^2}
  = \frac{C_F}{N_C} \cdot 5\,.
\end{equation}
The Mandelstam invariants are $\hat s = 4p_T^2$ and
$\hat t = \hat u = -2p_T^2$, and as a result the FEX channel has a
squared matrix element that is ten times larger than the FCR channel.

A second factor that is relevant is the partonic luminosity.
For the FEX channels, the incoming top is produced by an initial-state
$g \to t\bar t$ splitting, so ultimately the cross section is driven
by $gg$ and $g\Sigma$ luminosities, where $\Sigma$ is the sum of all
light (anti-)flavours.
The top-quark luminosity then involves a factor $\as \ln p_T/\mtop$,
which gives a smaller luminosity than either the $gg$ or
$q_i \bar q_i$ luminosities that were relevant for the FCR case.
Ultimately the larger matrix element compensates for the reduced
luminosities and the FEX process has a cross section that is
comparable to that for FCR.

A similar set of features emerges also for the GSP case.
Here the $\as \ln p_T/\mtop$ factor appears for the final-state
splitting rather than an initial state one.
It is straightforward to use massive splitting functions~\cite{Catani:2002hc} to
evaluate the leading-order probability
$\mathcal{P}_{g \to \ttbar}$ for $g \to \ttbar$ splitting with
the $\ttbar$ pair separated by distance $\Delta R_\ttbar < R$,
where
$\Delta R^2_\ttbar = (y_t - y_{\bar t})^2 + (\phi_t - \phi_\tbar)^2$
and $y_t$ and $\phi_t$ are respectively the rapidity and azimuth of
the top.
For a gluon transverse momentum of $p_T$, and with the conditions $p_T
R \gg \mtop$ and $R \ll 1$, the result is 
\logbook{}{../asymptotic-limit/g2QQbar-python-checks/g2QQbar.nb}
\begin{equation}
  \label{eq:g2ttbar-asymptotic}
  \mathcal{P}_{g \to \ttbar} =
  \frac{\as T_R}{2\pi} \frac23 \left(
    \ln \frac{p_{T,t}^2 R^2}{\mtop^2} - \frac{23}{6}
    \right)\,.
\end{equation}
In practice, the regime of $p_T = 1 \TeV$ is not sufficiently
asymptotic for this expression to hold, as one can see by substituting
$R=1$ and observing that the result is negative.
To obtain a more reliable estimate, we maintain the conditions
$p_T \gg \mtop$ and $R \ll 1$, but relax the constraint on
$p_T R/\mtop$.
The resulting expression is a little cumbersome,\footnote{
  Starting from the massive splitting function, one introduces $\mu =
  \mtop/(p_{T,t} \Delta R)$ (where $\Delta R$ is the separation between
  the top and anti-top), performs the logarithmic integral over $\mu$,
  and after rewriting the expression in terms $y$, which is the solution of $\mu^2 =
  -y^2/(1-y)^4$, integrates over $z$.
  One obtains
  \logbook{3478247}{../fabrizio-results/gsplit_analytic.nb}%
  \begin{multline}
    \label{eq:g2ttbar-semi-asymptotic-exact}
      \mathcal{P}_{g \to \ttbar} =
      \frac{\as T_R}{2\pi}  \left(
      -\frac{4 \left(y^4-9
        y^3+23 y^2-9 y+1\right)}{3
      (y-1)^3 \sqrt{y^2-6 y+1}} \tanh^{-1}\left(\frac{1-y}{\sqrt{y^2-6
          y+1}}\right)
    \right.   +  \\ \left.
    +\frac{2 \left(y^4+y^3+3 y^2+y+1\right) \log (y)}{3 (y-1)^3 (y+1)}
    -\frac{23}{9}
    \right)
  \end{multline}
  and the solution to take for $y$ is
  $y = 1 - \frac{i}{2\mu} + \frac{(1-i) \sqrt{4 \mu -i}}{2\sqrt{2}\mu}$.
}
but the following parametrisation
reproduces the correct result to better than $1\%$ for all values
of $p_T R/\mtop$
\logbook{}{See ../asymptotic-limit/g2QQbar-python-checks/wbb-checks.gp
  for the comparison of the Mathematica exact numerical result and our
  parametrisation}%
\logbook{}{See
 ../asymptotic-limit/alpgen-wbb-studies/wbb-checks.pdf
  for a comparison to alpgen for $W+b\bar b$, showing that the Mathematica result
  agrees well.}%
\begin{equation}
  \label{eq:g2ttbar-semi-asymptotic}
  \mathcal{P}_{g \to \ttbar} \simeq
  \frac{\as T_R}{2\pi}
  \frac{1}{3}\frac{\ln (1 + e^{4x - 23/3} + e^{2x}/10)}{1 -
    0.101 e^{-(x-2.2)^2/2.3}}\,,
  \qquad
  x = \ln \frac{p_{T,t} R}{\mtop}\,.
\end{equation}
In other small-$R$ calculations, corrections associated with finite
values of $R$ have often been found to go as $R^2$ with a small
coefficient~\cite{Dasgupta:2007wa}.
At this stage, there is some freedom in the $R$ value that we choose
in order to define the gluon-splitting.
Insofar as we are interested mainly in an order-of-magnitude estimate of
$g \to \ttbar$, we evaluate Eq.~(\ref{eq:g2ttbar-semi-asymptotic})
with $R=1$, ignoring potential $R^2$-suppressed corrections.
Substituting $p_t = 1\TeV$ and $\mtop = 173\GeV$ and
$\as(1\TeV) = 0.089$, this yields the result for
$\mathcal{P}_{g \to \ttbar}$ shown in Table~\ref{tab:channels}.
We see that, like FEX, the GSP topology is also comparable to the FCR
one.

Were we to consider significantly harder events (e.g.\ at a $100\TeV$
collider) or $b$-quarks instead of top quarks, the logarithmic factors
would start to become large, further enhancing the FEX and GSP
contributions relative to FCR.
This is consistent with earlier findings of large relative FEX and GSP
contributions to high-$p_T$ $b$-jet production~\cite{Banfi:2007gu}.

The analysis shown in Table~\ref{tab:channels} is not intended to give
precise predictions for the relative sizes of different topologies.
Nevertheless it shows that, despite their being suppressed by a power
of $\as$, the (NLO) FEX and GSP topologies are numerically comparable
to the LO FCR topology.\footnote{This finding is reminiscent of the
  observation of giant $K$-factors discussed for example in
  Ref.~\cite{Rubin:2010xp} for vector-boson plus jet production, though
  the $K$-factors in the $\ttbar$ case are less extreme.}
By framing the discussion in terms of an asymptotic limit where
$p_T \gg \mtop$, we avoided providing a rigorous definition of the
FCR, FEX and GSP topologies.
If one wishes to study actual events, whether in fixed-order QCD, or
at particle-level in experiments, a precise definition becomes
necessary. 
This will be the topic of Section~\ref{sec:parton-level-analysis}.

\subsection{Interplay between topologies and hardness characterisation
  variable}
\label{sec:topology-scale-interplay}

Before turning to detailed topology definitions, we discuss the
interplay between the topologies of Fig.~\ref{fig:classes} and the
event hardness variables of Table~\ref{tab:hard-scale-choices}.
While all three main topologies have comparable cross sections for
comparable hardness of the underlying $2\to2$ scattering, their
relative contributions to the differential distribution of some
specific event hardness variable depends significantly on the choice
of variable.
A key principle to remember in the discussion is that each of the
topologies has a cross section that falls steeply as a function of the
underlying $2\to 2$ transverse momentum $p_{T}^{2\to 2}$, say as
$\sim 1/(p_{T}^{2\to 2})^k$ with some positive power $k$.
Consider a specific value $V$ of a given hardness variable.
If $p_{T}^{2\to 2}$ is significantly larger than $V$ in some topology,
its contribution to the bin around $V$ will be suppressed relative to
another topology for which $p_{T}^{2\to 2}$ is comparable to $V$.
Equivalently, a topology where $V$ ends up being significantly smaller
than $p_{T}^{2\to 2}$ will be suppressed relative to a
topology where $V$ is similar to $p_{T}^{2\to 2}$.
On this basis we can work out which topologies will be relevant for
which hardness variable, and the conclusions are summarised in
Table~\ref{tab:hard-scale-topologies}.

\begin{table}
  \centering
  \begin{tabular}{ccccc}
    \toprule
    Hardness variable && FCR & FEX & GSP\\
    \midrule
    $p_{T}^{\text{top,had}}$
          && \checkmark & \checkmark & 
    \\
    $p_{T}^{\text{top,lep}}$
          && \checkmark & \checkmark & 
    \\
    $p_{T}^{\text{top,max}}$
          && \checkmark & \checkmark & 
    \\
    $p_{T}^{\text{top,min}}$
          && \checkmark & & 
    \\
    $p_{T}^{\text{top,avg}}$
          && \checkmark & & 
    \\
    \midrule
    $\frac12 H_T^{\ttbar}$  
          && \checkmark & & 
    \\
    $\frac12 \HT$ 
          && \checkmark & \checkmark & \checkmark
    \\
    $m_T^{J,\text{avg}}$
          && \checkmark & \checkmark & \checkmark
    \\
    \midrule
    $\frac12 m^{\ttbar}$
          && \checkmark & & 
    \\
    \midrule
    $p_{T}^{\ttbar}$
          && & \checkmark & \checkmark
    \\
    $p_T^{j_{\nott,1}}$
          && & \checkmark & \checkmark
    \\
    \bottomrule
  \end{tabular}
  \caption{
    Summary of the topologies expected to contribute dominantly to the
    distributions of different hardness variables, when these are
    large relative to $\mtop$.
    See text for details and Table~\ref{tab:hard-scale-choices} for
    definitions of the hardness variables.
  }
  \label{tab:hard-scale-topologies}
\end{table}

Specifically, we see that the first group of hardness variables in
Tables~\ref{tab:hard-scale-choices} and
\ref{tab:hard-scale-topologies}, the $p_T^\text{top}$ set of variables,
splits into two sub-groups.
The first three variables $p_{T}^{\text{top,had}}$,
$p_{T}^{\text{top,lep}}$, $p_{T}^{\text{top,max}}$ share the
characteristic that they can be commensurate with $p_{T}^{2\to 2}$ if
at least one of the two tops is hard.
Therefore we expect the distributions of these variables to receive
significant contributions from the FCR and FEX topologies,\footnote{In
  the asymptotically dominant limit where the softer of the two tops
  has negligible $p_T$ compared to the harder one, the FEX
  contribution for $p_{T}^{\text{top,had}}$ and
  $p_{T}^{\text{top,lep}}$ is half that for $p_{T}^{\text{top,max}}$.
  This will, however, not be unambiguously visible later when we
  compare FEX to FCR: the LO property that the FCR distributions
  of $p_{T}^{\text{top,had}}$, $p_{T}^{\text{top,lep}}$ and
  $p_{T}^{\text{top,max}}$ are identical is broken from NLO onwards,
  with the $p_{T}^{\text{top,max}}$ distribution being larger than the
  others. 
} but not from GSP (because neither of the tops carries the full $p_T$
of the underlying hard process).
In contrast, for $p_{T}^{\text{top,min}}$ and $p_{T}^{\text{top,avg}}$
to be commensurate with $p_{T}^{2\to 2}$, both tops need to be hard,
and so we expect significant contributions mainly from FCR.

The next set of variables in Tables~\ref{tab:hard-scale-choices} and
\ref{tab:hard-scale-topologies} also splits into two groups.
The $\frac12 H_T^{\ttbar}$ variable is commensurate with
$p_{T}^{2\to 2}$ only if both tops are hard, i.e.\ we expect
contributions mainly from FCR.
In contrast, $\frac12 \HT$ and $m_T^{J,\text{avg}}$ are always
commensurate with $p_{T}^{2\to 2}$, regardless of the underlying topology,
and so we expect contributions from FCR, FEX and GSP.

The $\frac12 m^{\ttbar}$ variable is special, as discussed in
Section~\ref{sec:hardness-scales} and
Appendix~\ref{sec:mttbar-beyond-LO}.
We do not expect significant FEX or GSP contributions associated with
NLO matrix elements that are larger than the LO one, and on that basis
would expect it to be dominated by FCR.
However if $\log m^\ttbar/\mtop$ becomes large enough, the discussion
of Appendix~\ref{sec:mttbar-beyond-LO} implies that the largest
log-enhanced contributions (e.g.\ $\as^3 \ln^3 m^\ttbar/\mtop$ terms)
would include FEX-like topologies.
At LHC energies, logarithms are not yet quite
large enough to override the main perturbative hierarchy and
accordingly we remain with the expectation that the $m^\ttbar$
distribution should mainly involve the FCR topology.

The last two variables that we consider are $p_{T}^{\ttbar}$ and
$p_T^{j_{\nott,1}}$, which are identical at NLO.
They are commensurate with $p_{T}^{2\to 2}$ only for the topologies
with a hard non-top jet, i.e.\ for the FEX and GSP topologies.

The issue of the relevant topologies is not the only aspect that
contributes to the size of the final cross sections for different
hardness variables.
We will comment on other aspects as they arise in the sections below.

\section{Parton-level (truth-top)  analysis}
\label{sec:parton-level-analysis}

If we are to explore the relevance of different topologies in actual
(simulated or experimentally observed) $\ttbar$ events, it becomes
necessary to develop techniques to identify the $\ttbar$ event topologies
based on the momenta of the top quarks and the other event particles.
Such techniques need to be applicable even outside the asymptotically
high-scale limit that formed the conceptual basis of the discussion in
Section~\ref{sec:theory}.

As a first step, we imagine a situation where we have full kinematic
information about the top and anti-top quarks and that we can separate
out all particles that do not stem from the $\ttbar$ decays.
In Section~\ref{sec:topology-from-tops} we outline a simple algorithm
to assign a classification of the topology for any given event.
Then in Section~\ref{sec:parton-level-results} we apply the algorithm
to simulated parton-level events and compare the results to the
expectations from Section~\ref{sec:theory}.

\subsection{Identification of topologies with identified tops}
\label{sec:topology-from-tops}

We consider a procedure for events with exactly one $\ttbar$ pair, and
follow a two-stage reconstruction procedure, set out as
Algorithm~\ref{alg:toppartons}.

\begin{algorithm}[h]
  \caption{Event analysis algorithm, given $t,\bar t$ partons, and other partons}
  \label{alg:toppartons}
  \renewcommand{\algorithmiccomment}[1]{[#1]}
  \begin{algorithmic}[1]
    \REQUIRE two undecayed tops, $t$, $\bar t$, and the set of partons
    not from top decay, $\{P_\nott\}$
    \STATE Cluster the non-top $\{P_\nott\}$ partons with the
    anti-$k_t$ algorithm~\cite{Cacciari:2008gp}, using a jet radius of
    $R=0.4$.
    \STATE Apply a transverse momentum
    threshold $p_{T,\min}$ to obtain the set of non-top $R=0.4$ jets,
    $\{j_\nott\}$, ordered in decreasing $p_T$. \label{alg:toppartons-smallR}
    \STATE Cluster the set $\{j_{\nott}, t, \bar t\}$ with a jet
    algorithm with radius of order $1$. Here we take the anti-$k_t$
    algorithm with radius $R_J=1$. \label{alg:toppartons:largeR}
    Refer to the resulting set of large-$R$ jets as $\{J\}$, which are
    sorted in order of decreasing $m_T^2 = p_T^2 + m^2$.
  \end{algorithmic}
\end{algorithm}

\noindent In the first stage we obtain $R=0.4$ jets from all objects other than
the top quarks.
This is the output of step~\ref{alg:toppartons-smallR} of the
algorithm.
The $R=0.4$ jet radius ensures that low-momentum particles from the
underlying event (UE) and pileup do not too significantly affect the
momenta of genuinely hard jets.
The $p_{T,\min}$ cut on the jets ensures that jets composed primarily
of low-momentum particles from UE and pileup do not significantly
affect variables that sum over multiple jets, such as $\HT$.
In this section we imagine a perfect detector, and apply no rapidity
acceptance cuts, neither on the top quarks nor on the jets.

Step~\ref{alg:toppartons:largeR} of the algorithm then clusters the
top quarks and the $R=0.4$ jets together using a jet radius of $1$.
When there are jets or top quarks at high transverse momentum, this
step effectively treats them democratically, reflecting a view that
the top quarks are akin to light partons, and should be included in
the clustering on the same footing as other partons.
Taking $R\simeq 1$ is the natural choice for separating initial-state
and final-state perturbative-QCD radiation~\cite{Soyez:2010rg}.
The use of $R=0.4$ jets (with a $p_{T,\text{min}}$ threshold) as the
input to the large-$R$ clustering ensures that the large-$R$ jets are
kept relatively free of underlying-event and pileup contamination, in
the same way as for observables such as $\HT$ that sum over multiple
jets. 
This clustering of smaller-radius jets into a larger radius system is
reminiscent of CMS's radiation recovery procedure in dijet resonance
searches~\cite{Chatrchyan:2011ns} and bears similarities also to the
use of filtering~\cite{Butterworth:2008sd} or
trimming~\cite{Krohn:2009th} with large-$R$ jets for resonance
reconstruction in Ref.~\cite{Cacciari:2008gd}.

Sorting the large-$R$ jets $\{J\}$ in order of decreasing $m_T^2$
ensures that for low-$p_T$ events, the first two large-$R$ jets always
contain the top quarks.
At large $p_T$, the difference between $m_T$ and $p_T$ ordering is
immaterial.

Given the output of Algorithm~\ref{alg:toppartons} for a specific
event, we then identify the topology as follows:
\begin{itemize}
\item If each of $J_1$ and $J_2$ contains one top (anti-)quark, we
  declare the event topology to be FCR.
\item If one of  $J_1$ and $J_2$ contains a single top, and the other
  does not contain a top, we declare the event topology to be FEX.
\item If one of $J_1$ and $J_2$ contains both tops, we declare the
  event topology to be GSP.
\item Otherwise, we define the event topology as ``other''.
\end{itemize}
One advantage of this simple approach is that it is straightforward to
implement and gives a definite answer about the topology of each
event.
%

\subsection{Results}
\label{sec:parton-level-results}

Let us now examine what happens when we analyse $t\bar t$ events
according to the procedure outlined so far.
We will consider events at a $pp$ centre-of-mass energy of
$\sqrt{s}=13\TeV$, simulated with the \texttt{hvq}
process~\cite{Frixione:2007nw} of POWHEG Box v2~\cite{Alioli:2010xd},
revision 3660, using the \texttt{PDF4LHC15\_nnlo\_mc} PDF
set~\cite{Butterworth:2015oua} via LHAPDF~\cite{Buckley:2014ana}, and
a top mass of $173\GeV$, interfaced to Pythia~8.240~\cite{Sjostrand:2014zea}, Monash13
tune~\cite{Skands:2014pea}, with multiple interactions turned off.
\logbook{}{FC checked scale used by POWHEG hvq is $m_T^{t}$ of the
  underlying Born configuration.}
All jet clustering is performed with FastJet~\cite{Cacciari:2011ma},
version 3.3.2.
In the future it would also be interesting to carry out similar
studies to NNLO
accuracy~\cite{Czakon:2015owf,Czakon:2016ckf,Catani:2019hip}, for
example taking advantage of recent developments in combining parton
showers and $\ttbar$ production at NNLO~\cite{Mazzitelli:2020jio}.

As inputs to Algorithm~\ref{alg:toppartons}, we take the truth tops
from the event record, together with all partons not coming from the
top decay.
Since the \texttt{hvq} process is NLO for $t\bar t$ production, we
expect to have FCR topologies accurate to NLO, and FEX and GSP
accurate to LO, while other topologies are at best generated by the
shower, so do not have any formal perturbative accuracy.
\logbook{fafcf960c}{See ttjet-analysis/compare-LO-v-NLO.pdf for
  comparison of hvq and $t\bar t+j$ POWHEG calculations}
For observables and topologies that start only at NLO ($\as^3$), we
have cross-checked the \texttt{hvq} results against the NLO
calculation for $\ttbar + \text{jet}$ (i.e.\ up to $\as^4$) in its
POWHEG implementation~\cite{Alioli:2011as} and found good agreement. 

In this section, even though we consider top quarks before their
decay, we will still label one of them as leptonic and the other as
hadronic.
Explicit cross sections will include the branching ratio for one top
to decay muonically and the other hadronically.

\begin{figure}
  \centering
  \includegraphics[width=0.33\textwidth,page=10]{plots/channels-parton9-truth.pdf}%
  \includegraphics[width=0.33\textwidth,page= 3]{plots/channels-parton9-truth.pdf}%
  \includegraphics[width=0.33\textwidth,page= 4]{plots/channels-parton9-truth.pdf}\\
  \includegraphics[width=0.33\textwidth,page= 2]{plots/channels-parton9-truth.pdf}%
  \includegraphics[width=0.33\textwidth,page= 9]{plots/channels-parton9-truth.pdf}%
  \includegraphics[width=0.33\textwidth,page=11]{plots/channels-parton9-truth.pdf}\\
  \includegraphics[width=0.33\textwidth,page= 5]{plots/channels-parton9-truth.pdf}%
  \includegraphics[width=0.33\textwidth,page= 6]{plots/channels-parton9-truth.pdf}%
  \includegraphics[width=0.33\textwidth,page= 1]{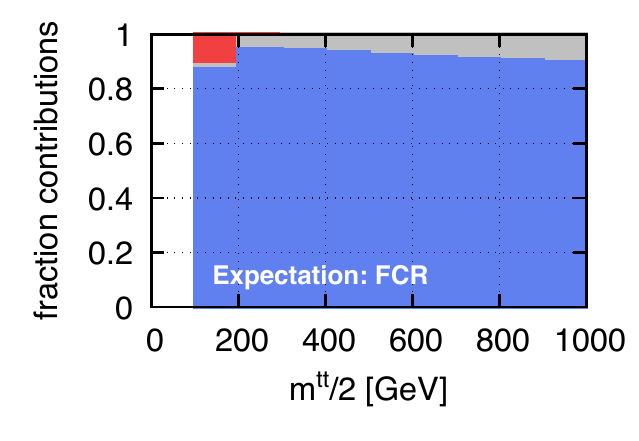}\\
  \includegraphics[width=0.33\textwidth,page= 7]{plots/channels-parton9-truth.pdf}%
  \includegraphics[width=0.33\textwidth,page= 8]{plots/channels-parton9-truth.pdf}%
  \includegraphics[width=0.33\textwidth,page=12]{plots/channels-parton9-truth.pdf}%
  \caption{Fractional contributions of the main top-production topologies
    (cf.\ Fig.~\ref{fig:classes}), as a function of the variable
    used to characterise the hardness of the event,
    cf.\ Table~\ref{tab:hard-scale-choices}.
    The expectations are those shown in Table~\ref{tab:hard-scale-topologies}.
  }
  \label{fig:truth-fractions-v-scale}
\end{figure}

For each of the event-hardness characterisation scales from
Table~\ref{tab:hard-scale-choices},
Figure~\ref{fig:truth-fractions-v-scale} shows the fractions of FCR,
FEX and GSP, as a function of the hardness scale (other topologies are
negligible).
The expectations from Table~\ref{tab:hard-scale-topologies} for the
dominant topologies at high scales are given on each plot.
Those expectations are all well borne out for sufficiently hard
events, $V \gtrsim 2\mtop$, and it striking to what extent different
event-hardness characterisation scale choices lead to very different
proportions of the various topologies.
In particular, the simple analysis of Table~\ref{tab:channels},
which suggested that all three topologies could potentially be of the
same size, is clearly reflected in the plots for the two hardness
scales that are insensitive to the particular topology, $\HT$ and
$m_T^{J,\text{avg}}$.
Within groups of observables that have the same dominant topologies
according to Table~\ref{tab:hard-scale-topologies}, there remain some
differences.
For example, $\HT$ has a larger FEX component than does
$m_T^{J,\text{avg}}$.
This is not surprising, because the $\HT$ variable receives a
contribution from the initial-state top that is present in FEX
topologies, while $m_T^{J,\text{avg}}$ does not.

\begin{figure}
  \centering
  \includegraphics[width=0.34\textwidth,page=1]{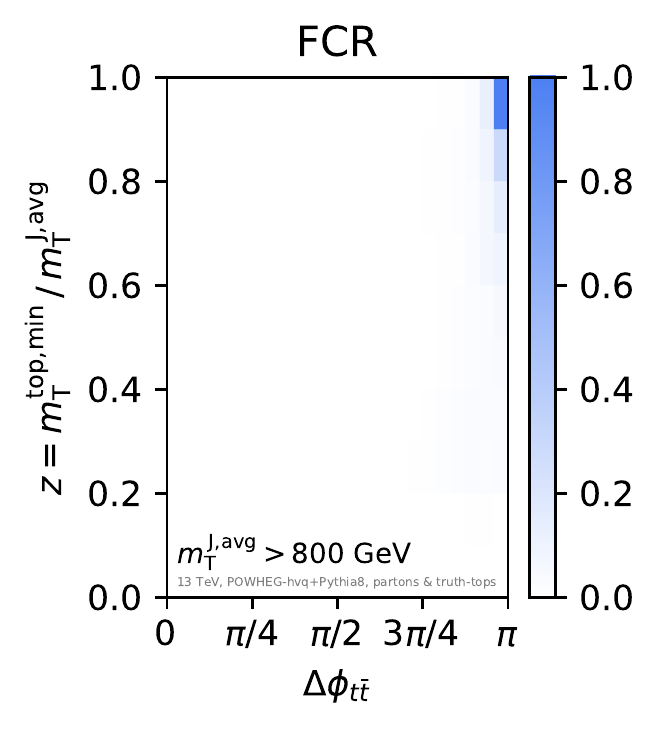}%
  \includegraphics[width=0.34\textwidth,page=2]{plots/2dpyplots.pdf}%
  \includegraphics[width=0.34\textwidth,page=3]{plots/2dpyplots.pdf}
  \caption{Kinematic distribution of the softer top, for events with
    large $m_T^{J,\text{avg}}$, in each of the three main topologies.
    The plots show $d\sigma/dzd\Delta\phi_{\ttbar}$, normalised to $1$
    at the maximum value of the histogram;
    $z = m_T^{\text{top,min}} / m_T^{J,\text{avg}}$ measures the
    hardness of the softer top relative to the underlying $2\to2$
    event hardness, and $\Delta\phi_{\ttbar}$ is the azimuthal
    distance between the two top quarks.
  }
  \label{fig:2dplots}
\end{figure}

To help visualise the kinematic features of the events in each of the
three main topologies, Fig.~\ref{fig:2dplots} shows the
two-dimensional distribution of $d\sigma/dzd\Delta\phi_{\ttbar}$ for
large $m_T^{J,\text{avg}}$ for each topology.
Here, $z = m_T^{\text{top,min}} / m_T^{J,\text{avg}}$ measures the
hardness of the softer top relative to the underlying $2\to2$ event
hardness, and $\Delta\phi_{\ttbar}$ is the azimuthal distance between
the two top quarks.
In FCR topologies we expect the softer top to balance against the
harder top ($z=1$) and to be back-to-back in azimuth
($\Delta \phi_{\ttbar} = \pi$).
In FEX topologies, because the softer top is produced through
initial-state radiation, we expect $z$ to take on predominantly low
values and $\Delta\phi_{\ttbar}$ to be spread out between $0$ and
$\pi$.
In GSP topologies, where the softer top is in the same jet as the
harder top, we expect $z \lesssim 1/2$ and
$\Delta\phi_{\ttbar} < R_J = 1$.
These features are broadly observed in the plots, though for the
finite values of $m_T^{J,\text{avg}}$ that we use, the exact limits on
$z$ are affected by the contributions of the top-quark mass to the
variables that enter its definition.

\begin{figure}
  \centering
  \includegraphics[width=0.48\textwidth,page=3]{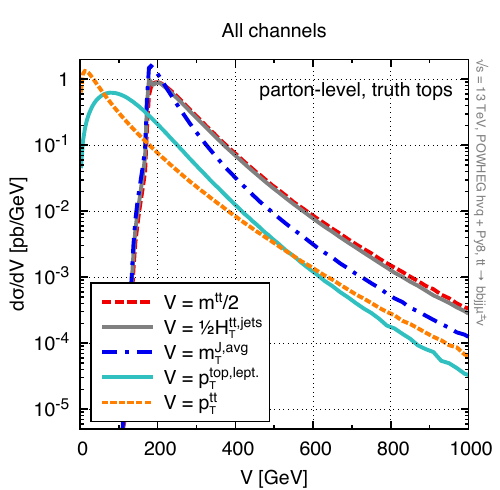}
  \includegraphics[width=0.48\textwidth,page=4]{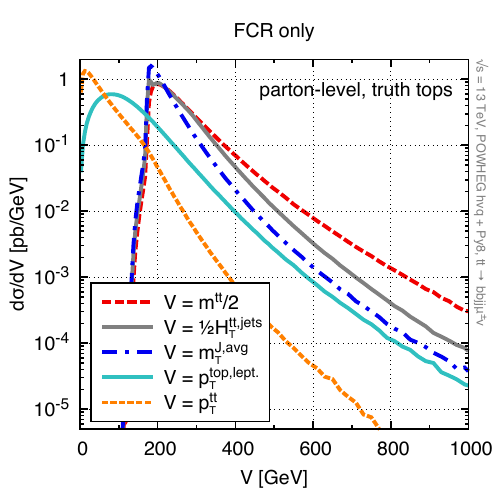}
  \caption{Differential cross sections as a function of a variety of
    scales used to characterise the event hardness ($V$), considering
    an illustrative subset of the scales from Table~\ref{tab:hard-scale-choices}.
    Left-hand plot: results summing over all topologies.
    Right-hand plot: results for just the FCR topology.
  }
  \label{fig:parton-cross-sections-4-V}
\end{figure}

Let us now apply the understanding that we have obtained to
investigate differential cross sections that are commonly studied
experimentally.
Fig.~\ref{fig:parton-cross-sections-4-V} shows differential cross sections
for a subset of observables, choosing at least one from each of the
groupings of Table~\ref{tab:hard-scale-choices}.
The left-hand plot shows the results without any topological
classification.
Among the features in the plot that is surprising at first sight is
that the $p_{T}^{\ttbar}$ distribution, which starts at $\as^3$, is
larger at high scales than the $p_{T}^{\text{top,lep}}$ distribution,
which starts at $\as^2$.
Based on the analysis of Section~\ref{sec:theory}, this is however not
a surprise, because of large FEX and GSP contributions to the
$p_T^{\ttbar}$ distribution.
If one considers only events with an FCR topology, as done in the
right-hand plot, the $p_{T}^{\ttbar}$ distribution ends up being
substantially suppressed relative to $p_{T}^{\text{top,lep}}$,
restoring faith in an analysis based on perturbation theory.

Another feature that becomes clearer when isolating the FCR topology
is the hierarchy between $\frac12 m^{\ttbar}$ on one hand and $\HT$,
$m_T^{J,\text{avg}}$ and $p_{T}^{\text{top,lep}}$ on the other.
A hierarchy is an expected consequence of the LO
$\log (m^{\ttbar} / \mtop) $ enhancement for the $\frac12 m^{\ttbar}$
distribution quoted in Eq.~(\ref{eq:sigma-mtt-LO}).
The other three observables are identical at LO, and free of any
$\log (m^{\ttbar} / \mtop) $ enhancement.
Yet in Fig.~\ref{fig:parton-cross-sections-4-V} (left) $\HT$ appears
to be almost identical to $\frac12 m^{\ttbar}$, and there is a clear
hierarchy among $\HT$, $m_T^{J,\text{avg}}$ and
$p_{T}^{\text{top,lep}}$.
If instead we examine Fig.~\ref{fig:parton-cross-sections-4-V}
(right), with just the FCR topologies, the pattern is closer to
the picture expected from LO: $\frac12 m^{\ttbar}$ is well above the
other observables, with a relative enhancement that increases towards
larger $\frac12 m^{\ttbar}$.
Meanwhile, $\HT$, $m_T^{J,\text{avg}}$ and $p_{T}^{\text{top,lep}}$
all show similar scaling at high momenta.
Remaining variations between them are straightforward to understand:
taking $m_T^{J,\text{avg}}$ as the reference, the $\HT$ variable
includes contributions from ISR radiation and so is larger, while
$p_{T}^{\text{top,lep}}$ is sensitive to the loss of radiation from
top fragmentation and so is smaller.

We have also checked the six other event-hardness scales from
Table~\ref{tab:hard-scale-choices} and the patterns observed are in
line with the analysis given above.

\begin{figure}
  \centering
  \includegraphics[width=0.5\textwidth,page=1]{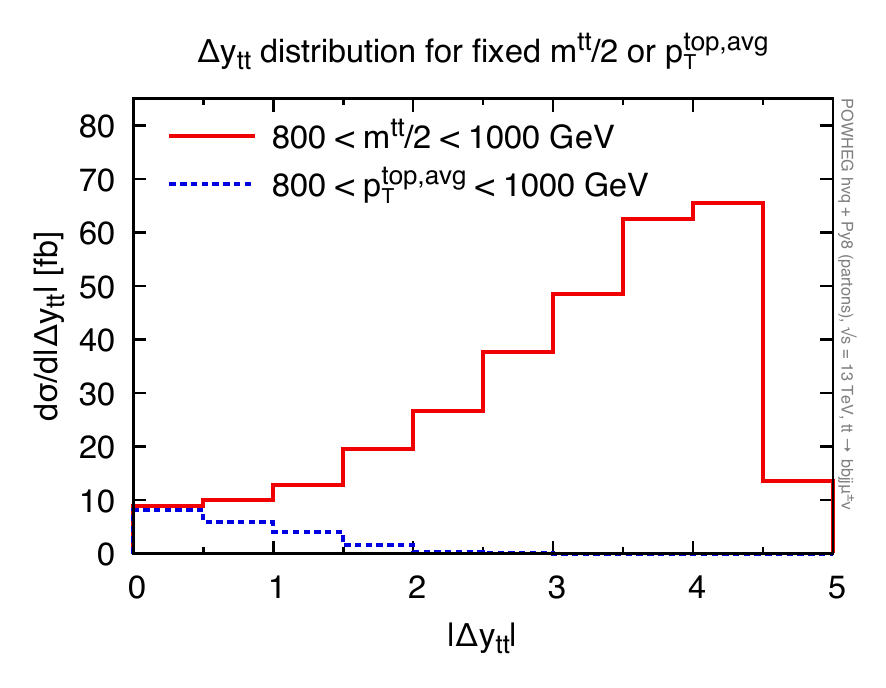}%
  \includegraphics[width=0.5\textwidth,page=3]{plots/delta-y-paper.pdf}
  \caption{
    Left: the distribution of $|\dytt|$ in for events where the $\ttbar$
    pair is in a given bin of either $m^{\ttbar} /2$
    or $p_{T}^{\text{top,avg}}$.
    Right: the average of the two top transverse momenta versus
    $|\dytt|$ in the given bin of $m^{\ttbar} /2$.
    We include all topologies in these results, keeping in mind that
    FCR is always the dominant contribution here.
  }
  \label{fig:dy-ttbar-parton}
\end{figure}

Our final comment of this section concerns the observation that the
$V=m^{\ttbar}/2$ distribution is $12{-}14$ times larger than the
$p_{T}^{\text{top,lep}}$ distribution.
\logbook{00874ee}{see plots/xsection-ratios.pdf}
A significant ratio is expected because of the $\log (m^{\ttbar} /
\mtop)$ enhancement that is present in Eq.~(\ref{eq:sigma-mtt-LO}),
associated the integral over $\dytt$ up to its kinematic boundary,
Eq.~(\ref{eq:dytt-max}).
Fig.~\ref{fig:dy-ttbar-parton} (left) shows the $\dytt$ distribution in a bin
of either the $m^{\ttbar} /2$ or $p_{T}^{\text{top,avg}}$ hardness scale.
In the lowest bin of $\dytt$, the results are independent of the
choice of hardness scale.
However at larger $\dytt$, the difference between the two histograms
is striking, with the $m^{\ttbar} /2$ case dominated by values of
$\dytt$ close to the kinematic limit, a consequence not just of the LO
distribution covering rapidities up to the kinematic boundary,
but also of further apparent logarithmic enhancements for large
$\dytt$ at NLO and beyond (cf.\ Appendix~\ref{sec:mttbar-beyond-LO}).
It is important to be aware that the events at large $\dytt$ involve
low transverse momenta for the top quarks.
This is illustrated in Fig.~\ref{fig:dy-ttbar-parton} (right), which
shows the average top-quark transverse momentum as a function of
$\dytt$ for the same bin of $m^{\ttbar}/2$ as shown on the left.
Close to the kinematic boundary of $\dytt$, where the cross section is
largest, the top quarks have transverse momenta of the order of
$\mtop$, which is to be expected given the basic kinematic relations
that hold at LO.

\section{Particle-level (reconstructed top) analysis}
\label{sec:hadronic-analysis}

In this section, we examine whether it is feasible to carry out an
analysis of $\ttbar$ topologies in actual collider events.
Standard experimental $\ttbar$ analyses take either a resolved
approach, or a boosted
approach~\cite{Larkoski:2017jix,Asquith:2018igt,Marzani:2019hun} to
identifying the top-quark decay products.
However, this strategy breaks down for the FEX topology because one
top is boosted, while the other may have only a moderate $p_T$.
It breaks down also for the GSP topology, because a single fat jet
contains both a top and an anti-top, and typical boosted-top tagging
algorithms are not designed to identify both tops.
Finally, we have seen that even in FCR topologies, a single bin of
$m^\ttbar/2 \gg \mtop$ receives contributions from tops at low as well
as high $p_T$, and so it is not sufficient to only apply boosted-top
tagging algorithms in measurements of the (high) $m^\ttbar/2$
distribution.

The strategy that we develop here is to adopt a resolved top
identification algorithm as our baseline, but to provide inputs such
that the algorithm continues to function even for tops at high $p_T$.
We will restrict our study to semi-leptonic $\ttbar$ events, and work
with the assumption that in a full experimental environment, lepton
tagging, double $b$-tagging and a missing-energy threshold would be
sufficient to reduce backgrounds to a manageable level.\footnote{We
  also ignore subtleties related to lepton identification in boosted
  top decays, see e.g.\ Ref.~\cite{Chatterjee:2019brg}.}

\subsection{Event analysis}
\label{sec:outline-ev-anl}

\begin{algorithm}[h]
  \caption{Event analysis algorithm at hadron (particle) level}
  \label{alg:event-analysis}
  \renewcommand{\algorithmiccomment}[1]{[#1]}
  \begin{algorithmic}[1]
    \REQUIRE at least one lepton (we require it to have a transverse
    momentum of at least $25\GeV$), missing transverse momentum and
    hadrons.
    \STATE Cluster the hadronic part of the event with the anti-$k_t$
    algorithm with $R=0.4$ and discard any jets below some $p_t$
    threshold, $p_{T,\min}$, as one would normally (we take $p_{T,\min}
    = 30\GeV$).
    \label{alg:event-analysis:cluster-j}
    \STATE Optionally, e.g.\ if subject to finite detector acceptance,
    exclude jets and leptons with an absolute rapidity beyond some
    $y_\text{max}$.
    The remaining set of jets is referred to as $\{j\}$ and the
    hadrons contained within that set of jets is $\{H\}$.
    \label{alg:event-analysis:max-rap}
    \STATE For each jet $j$, recluster its constituents with the exclusive longitudinally
    invariant ($R=1$) $k_t$ algorithm~\cite{Catani:1993hr} with a suitable
    $d_\text{cut}$ (we use $(20 \GeV)^2$), thus mapping the $R=0.4$
    jets $\{j\}$ to a declustered set $\{j_d\}$.
    One applies $b$-tagging to the $\{j_d\}$ (sub)jets to aid with the
    subsequent top identification.\label{alg:event-analysis:declust}

    \STATE Use a resolved top-tagging approach to identify the
    hadronic and leptonic top-quark candidates from the lepton(s) and
    from the jets $\{j_d\}$ obtained in
    step~\ref{alg:event-analysis:declust}.
    Here, we will adopt the algorithm outlined in
    Section~\ref{sec:reco}.
  
    \STATE Identify all particles from the set $\{H\}$ that do not belong to
    either of the top-quark candidates.
    Refer to this subset as $\{H_\nott\}$.
    Cluster the $\{H_\nott\}$ with the original jet definition
    (anti-$k_t$, $R=0.4$) and apply a transverse momentum threshold
    $p_{T,\min}$ to obtain the set of non-top $R=0.4$ jets,
    $\{j_\nott\}$, ordered in decreasing $p_T$.
    \label{alg:event-analysis:jnott}

    \STATE Apply step~\ref{alg:toppartons:largeR} of
    Algorithm~\ref{alg:toppartons} using $\{j_\nott\}$ and the
    reconstructed top and anti-top candidates as the inputs. 
  \end{algorithmic}
\end{algorithm}

The event analysis algorithm that we develop as a proof of concept is
given as Algorithm~\ref{alg:event-analysis}.
It is intended to function across the full range of top
transverse momenta with large LHC statistics, i.e.\ from low $p_T$ up
to $p_T\simeq 1\TeV$, and to be capable of reconstructing $\ttbar$
pairs that would lie in a single large-$R$ jet, thus addressing the
issues raised in the introduction to this section.
Some of the analysis steps involve a certain amount of choice.
When choosing between methods that are best at very high $p_T$ (very many
times the top mass) and methods that are simple, we have generally
chosen the latter.

Note that the declustering of step~\ref{alg:event-analysis:declust} is essential
when the hadronic top quark is at high $p_T$, because it resolves the
decay products even when they have been clustered into fewer than three $R=0.4$
jets $j$.\footnote{The declustering starts to have a significant
  impact on the reconstruction efficiency for
  $p_T^\text{top,had} \gtrsim 500 \GeV$.
  \logbook{8654a28}{../ttjet-analysis/2020-12-results/impact-ktrtdcut.pdf}
}
The approach that we use is similar to the early jet substructure work
of Seymour~\cite{Seymour:1993mx} and also coincides with the approach
adopted in the proposal~\cite{Apolinario:2017sob} to use tops to
characterise the time evolution of the heavy-ion medium.
If one were studying top quarks with very high $p_T$, it would
probably be better to develop an approach based on the
Cambridge/Aachen algorithm so as to reduce the high-$p_T$
top-tagging's sensitivity to the underlying event and pileup.
One might also wish to impose a constraint on the separation between
candidate top decay products that depends on the $p_T$ of the top
itself, similar to the variable-$R$ approach used in
Ref.~\cite{Kaplan:2008ie,Krohn:2009zg} or alternatively a
kinematically analogous fractional momentum cut on individual prongs,
as used in Soft Drop~\cite{Dasgupta:2013ihk,Larkoski:2014wba} and a
range of other taggers.

The choice to proceed via the $\{j_\nott\}$ set in step
\ref{alg:event-analysis:jnott} is motivated in particular if one
wishes to compare $R=0.4$ jet observables with purely resolved
measurements in the literature.

At very high $p_T$, instead of the $R_J=1$ anti-$k_T$ algorithm used in
step~\ref{alg:toppartons:largeR} of Algorithm~\ref{alg:toppartons}, it
might make more sense to adopt an algorithm such as
flavour-$k_T$~\cite{Banfi:2006hf,Banfi:2007gu} and possibly apply it directly to
the hadrons $\{H_\nott\}$ and tops, i.e.\ to the set
$\{H_\nott, t, \bar t\}$.
The flavour-$k_t$ algorithm suppresses the clustering of lone
soft-quarks within a hard jet, a situation which would contaminate the
flavour of a hard jet.\footnote{These configurations should be
  assigned to the ``other'' category of Fig.~\ref{fig:classes}, and
  this does not always occur with the anti-$k_T$ algorithm.
  The effects start only at order $\as^2 \ln p_T/\mtop$ relative to
  LO, and are practically negligible at the $p_T$ values that we study
  here, hence our choice to retain the simplicity of the anti-$k_T$
  algorithm.
  The effects are conceptually interesting when
  $ L = \ln p_T/m_\text{top} \gg 1$, because higher-order logarithms
  have a BFKL~\cite{Kuraev:1977fs,Balitsky:1978ic} structure, as
  pointed out by Marchesini and Mueller~\cite{Marchesini:2003nh}.
}

\subsection{Top reconstruction}
\label{sec:reco}

The top reconstruction that we use is a so-called ``resolved''
algorithm, i.e.\ one that takes advantage of the fact that the top
decay products should map to separate jets.
The declustering procedure in step~\ref{alg:event-analysis:declust} of
Algorithm~\ref{alg:event-analysis} helps ensure that this is true even
for high-$p_T$ tops.

There are many resolved top reconstruction algorithms in use for
semi-leptonic $\ttbar$ events, i.e.\ those with a lepton, missing
energy and jets, some of them $b$-tagged.
The procedure we adopt is largely based on the fiducial top definition
proposed in Ref.~\cite{CMS:2267573} and it makes use of invariant
masses in order to choose which jets to cluster together to form top
candidates.
Our version has one small modification concerning the neutrino
treatment.

First we reconstruct the neutrino momentum from the missing transverse
momentum $p_{T}^{\text{miss}}$ and the lepton 4-momentum
$p^{\text{lep}}$, setting
$\vec{p}_{T}^{\ \nu} = \vec{p}_{T}^{\text{ miss}}$ and determining
$p_{z}^{\nu}$ from the constraint
\begin{equation} 
  M(\nu_{l}+l) = m_{W}
\end{equation}
where $M(X)$ refers to the invariant mass of an object $X$ and
$m_{W}$ is the mass of the $W$ boson. 
Solving the resulting quadratic equation generally presents us with two 
solutions for $p_{z}^{\nu}$, the component of the neutrino's momentum along 
the beamline.
In cases where the two solutions are complex, we take their real part
as the physical $p_{z}^{\nu}$, and where both are real we take the
root with smaller $|p_{z}^{\nu}|$ (this is the one small point where
we differ from Ref.~\cite{CMS:2267573}, which takes the root with
larger $|p_{z}^{\nu}|$).
\logbook{2899be9cfc}{We run with `-neutrino-strategy ATLAS', which
  takes the smaller $|p_{z}^{\nu}|$ and this is the opposite of the
  CMS approach}

With this estimate of the kinematics of the leptonically decaying
W boson, we attempt to identify a subset of the 
reclustered (sub)jets $\{j_d\}$ with the remaining decay products.
Defining a semi-leptonic $\ttbar$ pair candidate by assigning one 
$b$-tagged jet to the leptonically decaying top quark candidate ($t_l$),
another to the hadronically decaying top candidate ($t_h$), and a pair of non 
$b$-tagged jets as the decay products of the hadronically decaying $W$ 
boson candidate ($W_h$), we calculate the quantity 
\begin{equation}
  K^2 = [M(t_{h}) - \mtop]^2 + [M(t_{l})-\mtop]^2 + [M(W_{h}) - m_{W}]^2
\end{equation}
for each combination of jets satisfying
$140 \GeV \leq M(t_{h/l}) \leq 190 \GeV$, where $\mtop = 173\GeV$ is
our top-quark mass choice (we do not place any direct requirements on
$M(W_{h})$).
If no such combination of jets exists then we deem the reconstruction 
to have failed, otherwise the $\ttbar$ candidate pair with the lowest $K^2$
is taken to most closely describe the kinematics of the full decay chain. 
%

\subsection{Validation of reconstruction performance}

\begin{figure}
  \centering
  \includegraphics[width=0.33\textwidth,page=1]{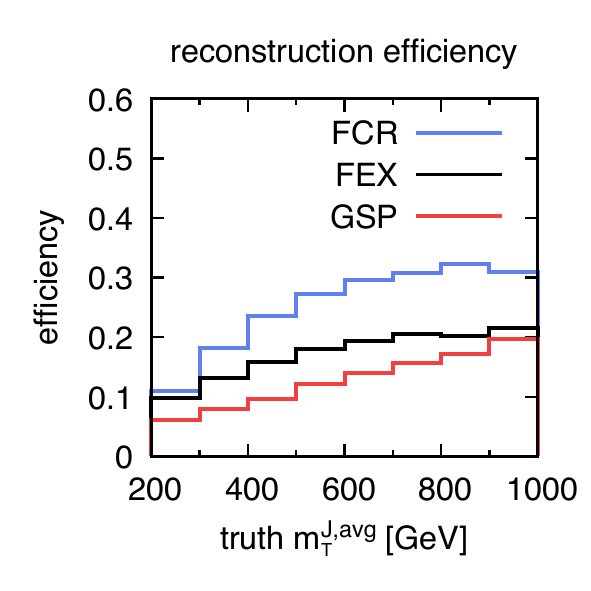}\hfill
  \includegraphics[width=0.33\textwidth,page=2]{plots/efficiency-purity.pdf}\hfill
  \includegraphics[width=0.33\textwidth,page=3]{plots/efficiency-purity.pdf}
  \caption{Tests of the efficiency and purity of the top
    reconstruction and topology identification procedures. See text
    for details.}
  \label{fig:purity-efficiency}
\end{figure}

To verify the performance of our reconstruction approach we consider
events at parton level, where we include the decays of the top quarks
and their subsequent showering.
The use of parton-level events here allows us to unambiguously
identify the source of each particle in the event, for example whether
it came from a $W$ decay.
We will show results as a function of $m_T^{J,\text{avg}}$, because
this event hardness characterisation scale yields events with a mix of
all three topologies, cf.\ Fig.~\ref{fig:truth-fractions-v-scale}.

Fig.~\ref{fig:purity-efficiency} (left) shows the overall efficiency
of the top reconstruction and topology identification.
Consider a topology that we wish to test, say FCR.
We identify all events where, using the Monte Carlo truth top quarks,
the topology was classified as FCR.
For a given bin of Monte Carlo truth $m_T^{J,\text{avg}}$, we then
determine the fraction of those events satisfying the following
conditions:
\begin{enumerate}
\item the event analysis algorithm of Section~\ref{sec:outline-ev-anl} and
  the top reconstruction algorithm of Section~\ref{sec:reco} (using the
  final leptons and partons) should successfully identify hadronic and
  leptonic top candidates;
\item additionally the reconstructed top candidates
  should predominantly contain the corresponding truth top decay
  products, specifically, the $b$ quarks should be correctly assigned in
  each candidate, and the two jets that make up the $W$ candidate
  should each have received at least $50\%$ of their $p_T$ from genuine
  $W$ decay products; \label{item:pure-top}
\item finally the event topology based on the reconstructed top
  quarks should also be FCR.
\end{enumerate}
One sees that the efficiency is about $10\%$ for low values of
$m_T^{J,\text{avg}}$, rising to $30\%$ at large $m_T^{J,\text{avg}}$,
with the FEX and GSP efficiencies being slightly lower than for FCR,
which is to be expected given that the FEX and GSP topologies are made
more difficult to reconstruct by the lower transverse momenta and/or
potential proximity of the top decay products.

We also verify the purity of the reconstruction, separating out the
study of the purity of the top reconstruction and of the topology
identification. 
Fig.~\ref{fig:purity-efficiency} (middle) shows the former.
For a given reconstructed topology, it shows the fraction of the
events in the given reconstructed $m_T^{J,\text{avg}}$ bin for which
the reconstructed top candidates predominantly contain the
corresponding truth top decay products (condition~\ref{item:pure-top}
above). 
The top purity is in the range $50{-}80\%$, increasing with
$m_T^{J,\text{avg}}$.
Fig.~\ref{fig:purity-efficiency} (right) shows the purity for the
topology identification.
Here we consider all events reconstructed as being in a given
topology, and examine the fraction for which the truth topology is the
same as the reconstructed one (irrespective of the whether the top
candidates match the truth ones).
This purity rapidly tends to $1$ with increasing $m_T^{J,\text{avg}}$.

Overall, the results of Fig.~\ref{fig:purity-efficiency} give us
confidence that the reconstruction approach proposed here can
be successfully applied to realistic events. 

\subsection{Results}
\label{sec:particle-level-results}

We close this article by repeating the main truth-level analyses of
Section~\ref{sec:parton-level-results} on hadron-level events (with
multi-parton interactions switched on), and imposing a realistic
detector rapidity acceptance, i.e.\ considering only jets and muons at
rapidities below $2.5$ in step~\ref{alg:event-analysis:max-rap} of
Algorithm~\ref{alg:event-analysis}. 

\begin{figure}
  \centering
  \includegraphics[width=0.33\textwidth,page=10]{plots/channels-hadron2-reco.pdf}%
  \includegraphics[width=0.33\textwidth,page= 3]{plots/channels-hadron2-reco.pdf}%
  \includegraphics[width=0.33\textwidth,page= 4]{plots/channels-hadron2-reco.pdf}\\
  \includegraphics[width=0.33\textwidth,page= 2]{plots/channels-hadron2-reco.pdf}%
  \includegraphics[width=0.33\textwidth,page= 9]{plots/channels-hadron2-reco.pdf}%
  \includegraphics[width=0.33\textwidth,page=11]{plots/channels-hadron2-reco.pdf}\\
  \includegraphics[width=0.33\textwidth,page= 5]{plots/channels-hadron2-reco.pdf}%
  \includegraphics[width=0.33\textwidth,page= 6]{plots/channels-hadron2-reco.pdf}%
  \includegraphics[width=0.33\textwidth,page= 1]{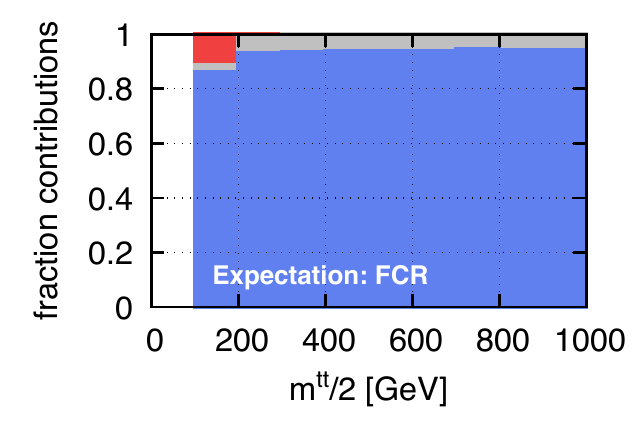}\\
  \includegraphics[width=0.33\textwidth,page= 7]{plots/channels-hadron2-reco.pdf}%
  \includegraphics[width=0.33\textwidth,page= 8]{plots/channels-hadron2-reco.pdf}%
  \includegraphics[width=0.33\textwidth,page=12]{plots/channels-hadron2-reco.pdf}%
  \caption{Fractional contributions of the main top-production topologies
    as a function of the variable
    used to characterise the hardness of the event.
    This is the analogue of
    Fig.~\ref{fig:truth-fractions-v-scale}, replacing Monte-Carlo
    truth top quarks with top-quark candidates reconstructed in
    particle-level events using the
    approach of sections \ref{sec:outline-ev-anl} and \ref{sec:reco}.
    A rapidity cut of $2.5$ has been applied to the muons and to the
    $R=0.4$ jets obtained in step~\ref{alg:event-analysis:cluster-j}
    of Algorithm~\ref{alg:event-analysis}.
  }
  \label{fig:hadron2-reco-fractions-v-scale}
\end{figure}

Fig.~\ref{fig:hadron2-reco-fractions-v-scale} is the analogue of
Fig.~\ref{fig:truth-fractions-v-scale} using top-quark candidates as
reconstructed from particle (hadron) level events.
The two sets of plots are strikingly similar, which should not be
surprising given the validation results shown above.
Where modest differences arise, these can be understood as a
consequence of the variations in reconstruction efficiencies across
different topologies.
For example one sees a slightly larger FCR contribution in the
hadron-level reconstructed $m_T^{J,\text{avg}}$ plot than in the
parton-level one, reflecting the higher efficiencies for FCR
reconstruction.
Similarly we have checked that the hadron-level reconstructed analogue of
Fig.~\ref{fig:2dplots} is close to the truth-level results.
\logbook{81abd43}{See plots/2dplots-hadron-reco.pdf}

\begin{figure}
  \centering
  \includegraphics[width=0.48\textwidth,page=3]{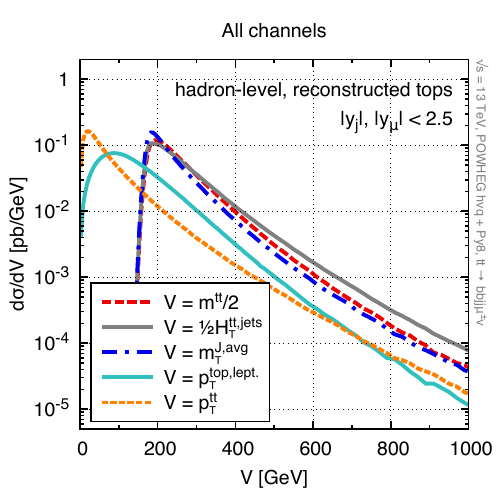}
  \includegraphics[width=0.48\textwidth,page=4]{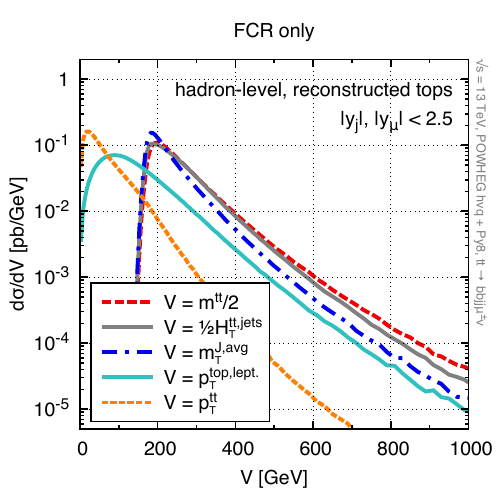}
  \caption{Hadron level results with reconstructed tops, for
    differential cross sections as a function of a selection of 
    scales used to characterise the event hardness ($V$), for a subset
    of such scales.
    Left-hand plot: results summing over all topologies.
    Right-hand plot: results for just the FCR topology.
    This plot is to be compared to
    Fig.~\ref{fig:parton-cross-sections-4-V}.
    The results here include a rapidity acceptance cut of $|y|<2.5$
    for jets and muons.
  }
  \label{fig:hadron-cross-sections-4-V}
\end{figure}

\begin{figure}
  \centering
  \includegraphics[width=0.58\textwidth,page=7]{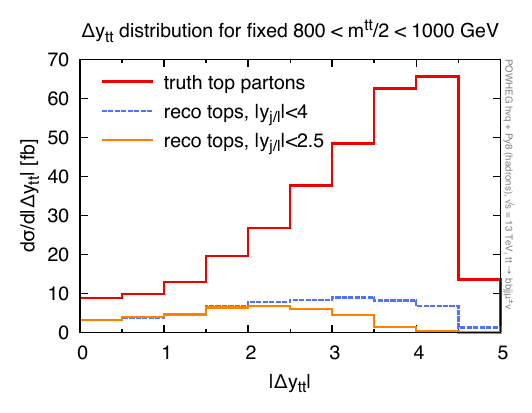}%
  \caption{
    Comparison of the truth (partonic-top) $\dytt$ distribution with the distribution
    obtained for fully reconstructed top quarks in hadron (particle) level
    events.
    The (truth or reconstructed)  $\ttbar$
    pair satisfies the constraint $800 < m^{\ttbar} /2 < 1000 \GeV$.
    The histograms include all topologies.
    \logbook{ec4d44a}{Compare pp.5 and 7 of this plot for parton v.\
      hadron} \logbook{81abd43}{and also
      plots/xsections-parton9-reco.pdf and
      plots/xsections-hadron9-reco.pdf}%
  }
  \label{fig:dy-ttbar-hadron}
\end{figure}

Fig.~\ref{fig:hadron-cross-sections-4-V} shows hadron-level
differential cross sections.
Broadly speaking the results are similar to those with truth tops in
Fig.~\ref{fig:parton-cross-sections-4-V}.
There is an overall reduction in the cross sections, which is to be
expected given the $10{-}30\%$ reconstruction efficiencies shown in
Fig.~\ref{fig:purity-efficiency}.

One additional striking difference is that the $m^\ttbar/2$
distribution no longer shows the strong enhancement relative to other
hardness scales at high values of $m^\ttbar/2$, for example being only
$4$ times larger than the $p_{T}^{\text{top,lep}}$ distribution at
around $1\TeV$, rather than $12{-}14$ times larger in
Fig.~\ref{fig:parton-cross-sections-4-V}. 
This can be understood from Fig.~\ref{fig:dy-ttbar-hadron}, which
compares results for $\dytt$ at large $m^\ttbar/2$ using truth tops
and using tops as reconstructed from the final particles (in this case final
partons).\footnote{For a related measurement see
  Ref.~\cite{Sirunyan:2019zvx}.
  The measurement is for a difference of top pseudorapidities,
  $\Delta \eta^{\ttbar}$, in a given bin of $m^\ttbar$, rather than a
  difference of rapidities, $\dytt$.
  The features should broadly speaking be similar, though it is useful
  to be aware that $\Delta \eta^{\ttbar}$ is not invariant under
  longitudinal boosts, while $\dytt$ is.
  Furthermore, for $p_T^\text{top} \lesssim \mtop$ the pseudorapidity
  correlates less well with the pseudorapidities of the top decay
  products than does the rapidity of the top. 
}
The reconstruction procedure has the largest impact at large values of
$\dytt$, where the top quarks have relatively low $p_T$ (which tends
to reduce reconstruction efficiencies) and where additionally some of
their decay products may fall beyond the rapidity acceptance.
Fig.~\ref{fig:dy-ttbar-hadron} shows results for a current LHC
acceptance of $|y|<2.5$ and for an HL-LHC type acceptance of $|y|<4$.
The latter is almost identical to results with full rapidity
acceptance.
\logbook{}{See other plots in the same file}%
Since the enhancement of the truth $m^\ttbar/2$ distribution is
precisely due to the contributions from large $\dytt$, it is now evident why
that enhancement is reduced with the reconstructed tops in
Fig.~\ref{fig:hadron-cross-sections-4-V}.
This suggests that for any kind of precision study with the
$m^\ttbar/2$ variable it would be wise to apply an upper limit on
$|\dytt|$, e.g.\ $|\dytt| < 2$.
On one hand, within this region, Fig.~\ref{fig:dy-ttbar-hadron} shows
that measurements would then be largely unaffected by the experimental
rapidity acceptance cuts on the input jets and leptons.
On the other hand, the $|\dytt| < 2$ rapidity window is still large
enough, for example, to exploit the different $\dytt$ dependencies of
the $q\bar q$ and $gg$-induced channels to separately constrain gluon
and (anti)-quark PDFs, cf.\ Fig.~\ref{fig:rap-structure} of
Appendix~\ref{sec:LO-notes-mtt}.
%

\section{Conclusions}
\label{sec:conclusions}

The core observation of this paper is that energetic $\ttbar$
production involves far more than the simple leading-order
back-to-back topology (flavour creation, FCR).
If one selects events with a large momentum transfer and a $\ttbar$
pair, one finds roughly equal contributions from the FCR topology, and
from each of the two topologies that start only at NLO: events with an
initial-state $g\to \ttbar$ splitting (flavour excitation, FEX)
followed by a hard scattering of one of the tops, and events with a
non-top hard scattering that is followed by a final state
$g\to \ttbar$ splitting (gluon splitting, GSP).
The $\HT$ and $m_T^{J,\text{avg}}$ panels of
Fig.~\ref{fig:truth-fractions-v-scale} illustrate that all three of
these topologies have similar cross sections.
The reason for this surprising pattern is that the underlying hard
$2\to2$ scattering channels that dominate at NLO involve
$t$-channel gluon exchange, and so have squared matrix elements that are an
order of magnitude larger than for the LO scattering channels, which
involves either an $s$-channel gluon exchange or a $t$-channel quark
exchange.
That enhancement numerically compensates for the extra factor of $\as$
that appears from the $g \to \ttbar$ splitting in the FEX and GSP
channels, as can be seen from Table~\ref{tab:channels}.
The specific mix of topologies depends critically on the choice of
observable used to characterise the hardness of $\ttbar$ events, as
anticipated in Section~\ref{sec:topology-scale-interplay} and verified 
Fig.~\ref{fig:truth-fractions-v-scale}.

An awareness of the role played by different topologies is important
in experimental measurements.
Identifying the top quarks in FEX and GSP topologies brings additional
challenges as compared to the FCR topology, for example because the
two tops may have significantly different transverse momenta, or
because they may end up within a single jet.
This requires the use of top reconstruction techniques that transcend
the usual resolved versus boosted paradigm, and we proposed a suitable
fiducial-top definition in Section~\ref{sec:hadronic-analysis}, which
combines the approaches of Refs.~\cite{Apolinario:2017sob} and
\cite{CMS:2267573}.
The use of techniques that can successfully identify the $\ttbar$ pair
in all topologies is especially critical for measurements that correct
to the level of top-partons: if a measurement is blind to a certain
topology, then the contribution from that topology to the final cross
section risks being estimated entirely from simulation rather than
data.%
\footnote{
  Similar issues with top reconstruction apply for the $m^\ttbar$
  distribution, independently of any questions of topology, because at
  large $m^\ttbar$ the $\ttbar$ quarks themselves may both have low
  $p_T$ or high $p_T$, depending on their rapidity separation.
  Given that large rapidity separations tend to be difficult to
  reconstruct, because of finite detector acceptances, cf.\
  Fig.~\ref{fig:dy-ttbar-hadron}, it might be advisable for future
  measurements of the $m^\ttbar$ distribution to include a cut on
  $\dytt$, e.g.\ $|\dytt| < 2$, as discussed at the end of
  Section~\ref{sec:particle-level-results}. 
}

As we saw in Sections~\ref{sec:parton-level-analysis} and
\ref{sec:hadronic-analysis} it is  possible to provide
classifications of the topology of individual events.
Given that these topologies involve different PDFs and different
underlying $2\to2$ hard-scattering processes, separating out those
topologies can help in extracting the most physics information from
the data.
This is potentially relevant in any use of energetic $\ttbar$
production for precision physics, whether PDF fits, EFT studies,
searches for small direct BSM, or validation of simulation tools.
A separation by topology would also seem wise when studying the
order-by-order convergence of perturbative predictions.
We therefore encourage future measurements and theoretical studies to
further explore the rich topological structure of energetic $\ttbar$
production.

One consideration that we have not explored in any depth is that of
enhancements of perturbative contributions by logarithms of the
hardness scale divided by the top mass (e.g.\
FONLL~\cite{Cacciari:1998it,Cacciari:2018qlp}, BFKL in both
$t$-channel
quark~\cite{Sen:1982xv,Fadin:1976nw,Fadin:1977jr,Bogdan:2006af} and
gluon exchange~\cite{Kuraev:1977fs,Balitsky:1978ic}, BFKL in EW
exchange~\cite{Bartels:2006kr}, double
logarithms~\cite{Mueller:1985zp,Mangano:1992qq,Seymour:1994bz,Seymour:1994ca}
and BFKL logarithms~\cite{Marchesini:2003nh,Marchesini:2004ne,Marchesini:2015ica}
in final state fragmentation, and double-log small-$x$ non-singlet
enhancements~\cite{Kirschner:1983di,Ermolaev:1995fx}).
At LHC energies, such logarithms only start to become relevant for
$\ttbar$ production (cf.\ Eq.~(\ref{eq:g2ttbar-asymptotic})).
However, theoretically, the breadth of different classes of
logarithmic enhancement would make for a compelling study.
Such a study would probably be called for at higher collider energies
(e.g.\ the $100\TeV$ of the FCC-hh), and could
potentially also be of interest for $b\bar b$ production at the
HL-LHC.

\section*{Acknowledgements}

We are grateful to the authors of Ref.~\cite{Mazzitelli:2020jio} for
supplying us with the MiNNLO event sample that we used for
cross-checks in Appendix~\ref{sec:mttbar-beyond-LO}, to Pier Monni,
Paolo Nason and Giulia Zanderighi for comments on the manuscript and
to Lucian Harland-Lang and Gregory Soyez for helpful discussions.
This work was supported by the Science and Technology Facilities
Council (STFC) under grants ST/P000770/1 (FD) and ST/T000864/1 (FC and
GPS),
by a Royal Society Research Professorship
(RP$\backslash$R1$\backslash$180112)
and by the European Research Council (ERC) under the European Union’s
Horizon 2020 research and innovation programme, grant agreements
804394 (HipQCD, FC) and 788223 (PanScales, GPS).
RWM wishes to acknowledge the Rudolf Peierls Centre for Theoretical
Physics for support in the context of the Undergraduate Research
Opportunities Programme.

\appendix
\section{Leading-order distributions}
\label{sec:LO-notes}

For reference, we present and comment on some analytic formulas for
leading-order $t\bar t$ cross sections, in a limit where one 
kinematic variable is much larger than the top mass.
The results are essentially textbook level, and can be
straightforwardly be derived from matrix elements to be found, e.g., in
Ref.~\cite{Ellis:1991qj}.
They help provide some of the background to
Section~\ref{sec:hardness-scales}.

\subsection{Distributions differential in $m_\ttbar$}
\label{sec:LO-notes-mtt}

While in the main text we have used $m^\ttbar/2$ for distributions,
to keep the notation more compact, here we use the $\ttbar$ mass
rather than half the mass, and write it as $m_\ttbar$.
We also write $m_t \equiv \mtop$. 
We consider the region $m_\ttbar \gg \mtop$.
We start with a distribution that is double differential in
$m_\ttbar$ and the rapidity between the top and anti-top quarks,
$\Delta y_{\ttbar} = y_t - y_{\bar t}$.
\begin{equation}
  \label{eq:5}
  \frac{d\sigma}{dm_\ttbar^2 d\dytt} \simeq
  \frac{\as^2 \pi}{m_\ttbar^4} \left[
    \frac{(8 \cosh \dytt -
      1)\cosh \dytt}{48 (1 + \cosh \dytt)^2} \mathcal{L}_{gg}(m_\ttbar^2/s)
    +
    \frac29 \frac{\cosh \dytt}{(1 + \cosh \dytt)^2}
    \mathcal{L}_{q\bar q}(m_\ttbar^2/s)
  \right].
\end{equation}
The partonic luminosities are defined as
\begin{equation}
  \label{eq:lumi-def}
  \mathcal{L}_{ij}(x) = (2 - \delta_{ij})\int_x^1 \frac{dz}{z} zf_i(z) \frac{x}{z}f_j(x/z)\,,
\end{equation}
where $zf_i(z)$ is the distribution of partons of flavour $i$ and momentum
fraction $z$ and $\mathcal{L}_{q\bar q}$ sums over $q\bar q$ flavours.
In Eq.~(\ref{eq:5}), we have neglected corrections of the form
$\mtop^2/(m_T^{t})^2$.
For $\dytt \lesssim 1$, the condition $m_\ttbar \gg m_t$ implies large $m_T^{t}
\gg m_t$ and so they can be neglected.
For large $\dytt$, one can have $m_T^{t} \sim m_t$, however one can
verify that the
relative contribution of the $m_t^2/(m_T^{t})^2$ terms is suppressed
by a power of $\cosh \dytt$.

\begin{figure}
  \centering
  \includegraphics[width=0.48\textwidth,page=1]{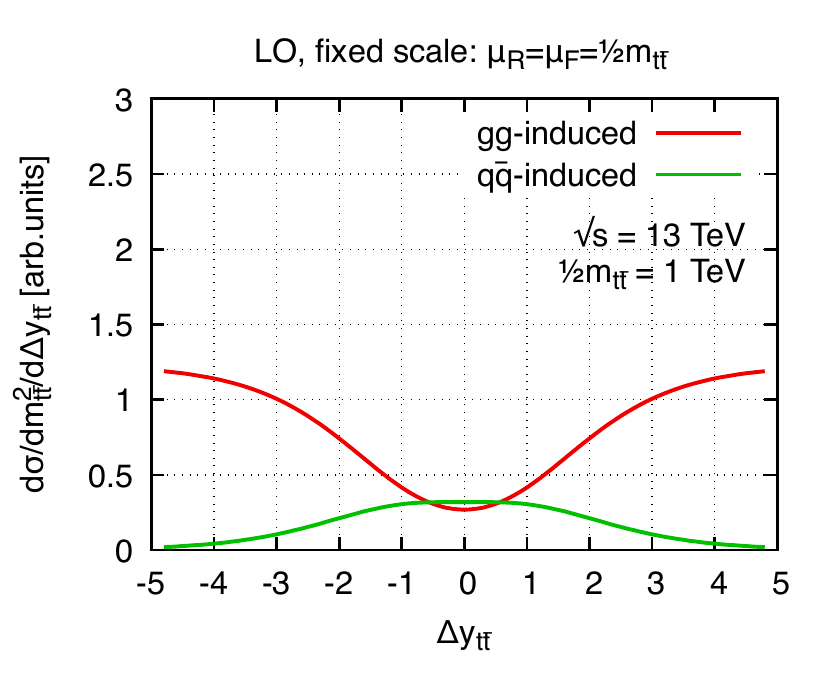}
  \includegraphics[width=0.48\textwidth,page=2]{plots/dytt-fixed-mtt-paper.pdf}
  \caption{Asymptotic leading-order forms for the $\dytt$ distribution at fixed
    large $m_\ttbar$, as given in Eq.~(\ref{eq:5}), applied to the case of
    $\sqrt{s}=13\TeV$ $pp$ collisions and $m_\ttbar = 2\TeV$.
    The left-hand plot uses fixed renormalisation and factorisation
    scales, which is physically 
    inappropriate but illustrates the key analytical features of
    the structure of Eq.~(\ref{eq:5}).
    The right-hand plot uses a physically motivated scale choice, of
    the order of the momentum transfers involved in the process.
  }
  \label{fig:rap-structure}
\end{figure}

In Eq.~(\ref{eq:5}), for large $\dytt$, the term proportional to the
gluon luminosity becomes independent of $\dytt$, i.e.\ one obtains a
flat distribution in $\dytt$.
In contrast, the term proportional to the quark-antiquark luminosity
vanishes.
This difference in behaviour is a consequence of the fact that the
$gg$ channel includes a $t$-channel quark-exchange diagram, whereas
the $q\bar q$ channel only involves $s$-channel exchanges.

The result in Eq.~(\ref{eq:5}) is valid up to the kinematic limit
given in Eq.~(\ref{eq:dytt-max}),
$\dytt^{\max} \simeq \ln \frac{m_\ttbar^2}{m_t^2}$.
It is illustrated for fixed scale in the coupling and PDFs in
Fig.~\ref{fig:rap-structure} (left), separated into the $gg$ and
$q\bar q$-induced components.
However, it is physically inappropriate to use a fixed
scale, even for a single value of $m_\ttbar$, because different
rapidities involve substantially different transverse momenta, and
associated momentum transfers.
The impact of using a physically motivated scale choice,
$\mu_R^2 = \mu_F^2 = (H_T^{\ttbar}/2)^2 = m_\ttbar^2 /
(2(1+\cosh\dytt))$, is shown in the right-hand plot.\footnote{
For more complex events, one may choose to generalise this to $m_T^{J,\text{avg}}$ or
$\HT/2$ of Table~\ref{tab:hard-scale-choices}.}
This choice has a major impact on the shape of the distribution, with
the plateaus in the $gg$-induced distribution acquiring a strong
quasi-linear dependence on $\dytt$.
This dependence arises from the scaling violations in the coupling and
PDF, a consequence of $\ln \mu^2 \simeq \ln m_\ttbar^2 - \dytt$.
The precise slope depends on the $x$ values being probed in the
PDF. 

The significant difference in $\dytt$ dependence for $q\bar q$ and
$gg$-induced production has the potential to provide a valuable handle
separately for the gluon and quark parton distributions.

We can also integrate over $\dytt$ to obtain the single-differential
distribution,
\begin{equation}
  \label{eq:4}
  \frac{d\sigma}{dm_\ttbar^2} = \frac{\alpha_s^2 \pi}{m_\ttbar^4}
  \left[
    \left(\frac13 \ln \frac{m_\ttbar^2}{m_t^2} - \frac7{12} \right)
   \mathcal{L}_{gg}
    +
    \frac{8}{27}\, \mathcal{L}_{q\bar q}
  \right],
\end{equation}
quoted in Section~\ref{sec:hardness-scales} as
Eq.~(\ref{eq:sigma-mtt-LO}).
Again we have neglected corrections that are suppressed in
our kinematic region.
The result is obtained for a fixed scale and, as discussed above, this
is a physically inappropriate choice. 
Nevertheless, it is instructive to have the analytical result in this
limit, because it reveals a $\ln \frac{m_\ttbar^2}{m_t^2}$ enhancement
of the gluon-induced contribution.
%

\subsection{Distributions differential in the top transverse momentum}
\label{sec:LO-notes-pt}

For large $\pTt \gg m_t$, the leading order top-quark distribution
doubly-differential in $\pTt$ and $\dytt$ is given by
\begin{multline}
  \label{eq:7}
  \frac{d\sigma}{d\pTt^2 d\dytt}
  \simeq \frac{\as^2 \pi}{4\pTt^4} \left[
    \frac{(8 \cosh \dytt -
      1)\cosh \dytt}{24 (1 + \cosh \dytt)^3}
    \mathcal{L}_{gg}\left(2(1 + \cosh \dytt)\pTt^2/s\right)
    +
  \right.\\  \left.
    +
    \frac49 \frac{\cosh \dytt}{(1 + \cosh \dytt)^3}
    \mathcal{L}_{q\bar q}\left(2(1 + \cosh \dytt)\pTt^2/s\right)
  \right]\,,
\end{multline}
neglecting terms relatively suppressed by powers of $m_t^2 / \pTt^2$.
The structure here is very similar to that of Eq.~(\ref{eq:5}), and
indeed it is a trivial rewriting of that result since at LO any
combination of $\pTt$ and $\dytt$ can be mapped to a combination of
$m_\ttbar$ and $\dytt$.
Still, some features change: we have an overall factor of
$1/\pTt^4$ rather than $1/m_\ttbar^4$;
inside the outer square bracket the denominators have three powers of
$(1+\cosh\dytt)$ rather than two.
These are trivial consequences of the LO relation between $m_\ttbar$
and $\pTt$, and of the Jacobian transformation in the differential
cross section.
Importantly, for a given bin of $\pTt$ the luminosities are evaluated
at a mass that now depends on the rapidity separation, whereas in a
given bin of $m_\ttbar$ they didn't.

\begin{figure}
  \centering
  \includegraphics[width=0.48\textwidth,page=1]{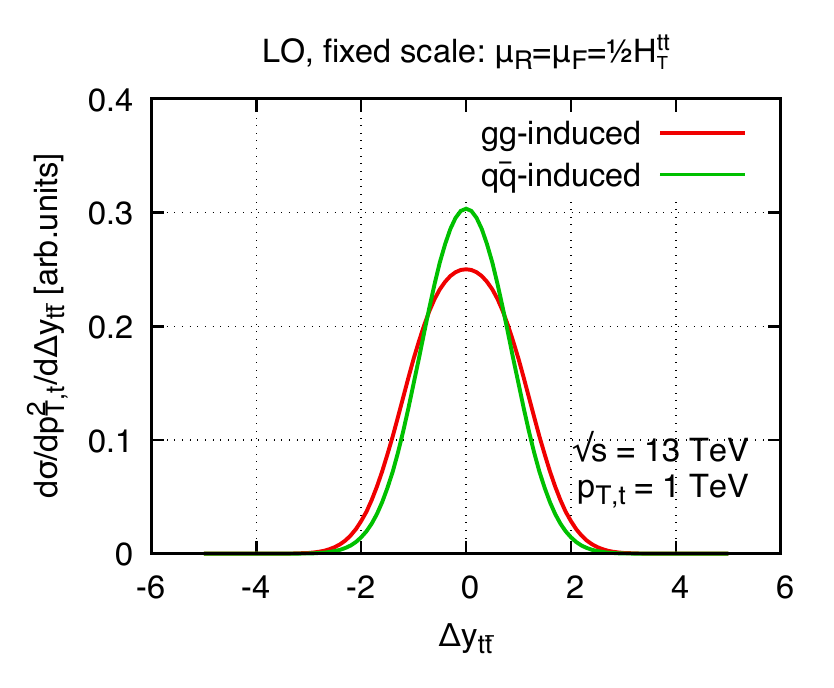}
  \includegraphics[width=0.48\textwidth,page=1]{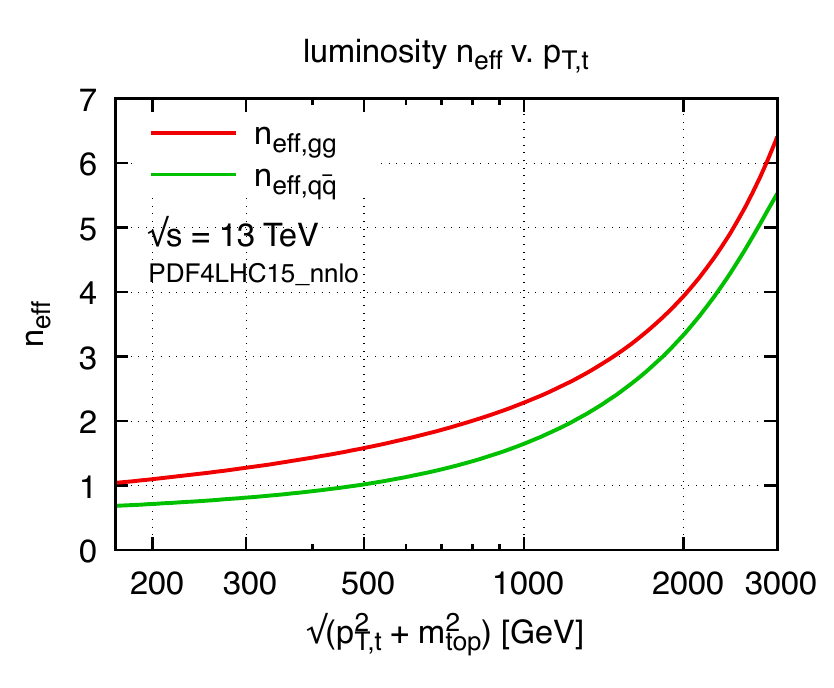}
  \caption{Left: Asymptotic leading-order forms for the $\dytt$ distribution at fixed
    large $\pTt$, as given in Eq.~(\ref{eq:7}), applied to the case of
    $\sqrt{s}=13\TeV$ $pp$ collisions and $\pTt = 1\TeV$.
    Right: effective power-law dependence, $n_\text{eff}$ of the
    partonic luminosities, as a function of $H_T^{\ttbar}/2=
    \sqrt{\smash[b]{p_{T,t}^2 + \mtop^2}}$.
    See text for details. 
  }
  \label{fig:rap-structure-fixed-pTt}
\end{figure}

The distributions for the rapidity difference are illustrated in
Fig.~\ref{fig:rap-structure-fixed-pTt}.
While the $gg$-induced term still has a broader $\dytt$ distribution
than the $q\bar q$-induced term, the difference in shapes is much
smaller than for fixed $m_\ttbar$.
In particular, both distributions are now concentrated around
$\dytt = 0$ and quite strongly peaked there: this is partly because of
the extra power of $(1+\cosh\dytt)$, and partly because the partonic
luminosities drop off rapidly with increasing rapidity separation.
The kinematic limit at fixed $\pTt$ is reached when
$2(m_t^2 + \pTt^2)(1+\cosh\dytt) = s$, giving
\begin{equation}
  \dytt^{\max} | \pTt = \cosh^{-1}\left(\frac{s}{2(m_t^2 + \pTt^2)}-1\right)
  \simeq \ln \frac{s}{\pTt^2}\,,
\end{equation}
where $s$ is the proton--proton centre-of-mass energy and the
approximation in the right-hand answer is valid for
$m_t^2 \ll \pTt^2 \ll s$.

The dependence of the partonic luminosities in Eq.~(\ref{eq:7}) on the
rapidity separation means that one cannot analytically integrate over
$\dytt$ to obtain $d\sigma /d\pTt$.
However, for the purpose of understanding the key structural elements
we make the approximation that the luminosity scales with the
system mass as a power law, 
\begin{equation}
  \label{eq:8}
  \mathcal{L}(m^2) \propto \left(\frac{1}{m^2}\right)^{n_\text{eff}}\,.
\end{equation}
Equivalently, one may define
\begin{equation}
  \label{eq:9}
  n_\text{eff} = - \frac{d\ln \mathcal{L}(m^2)}{d\ln m^2}\,.
\end{equation}
The effective power $n_\text{eff}$ is shown in
Fig.~\ref{fig:rap-structure-fixed-pTt} (right) as a function of the
top transverse mass (or equivalently, half the system mass for $\dytt
= 0$).
With this approximation, one can integrate Eq.~(\ref{eq:7}) over
$\dytt$ to obtain
\begin{equation}
  \label{eq:10}
  \frac{d\sigma}{d\pTt^2}
  \simeq \frac{\as^2 \pi}{4\pTt^4} \left[
    f_{gg}(n_{\text{eff},gg}) \mathcal{L}_{gg}(4\pTt^2/s)
    +
    f_{q\bar q}(n_{\text{eff},gg}) \mathcal{L}_{q\bar q}(4\pTt^2/s)
  \right]
\end{equation}
where $f_{gg}(n_{\text{eff},gg})$ and $f_{q\bar q}(n_{\text{eff},gg})$
are purely numerical factors, which can be expressed in terms of
hypergeometric functions of $n_\text{eff}$, but are perhaps more
usefully shown for specific values of $n_{\text{eff}}$, cf.\
Table~\ref{tab:fxx-neff}.
\begin{table}
  \centering
  \begin{tabular}{ccccc}
    \toprule
    $n_{\text{eff}}$ & 1 & 2  & 3 & 4 \\
    \midrule
    $f_{gg}(n_{\text{eff}})$ & $\frac{16}{105}$
                             & $\frac{34}{315}$
                             & $\frac{128}{1485}$
                             & $\frac{3296}{45045}$\\[5pt]
    $f_{q\bar q}(n_{\text{eff}})$
                             & $\frac{128}{945}$
                             & $\frac{64}{567}$
                             & $\frac{1024}{10395}$
                             & $\frac{1024}{11583}$
    \\[3pt]
    \bottomrule
  \end{tabular}
  \caption{The coefficients that appear in Eq.~(\ref{eq:10}) for
    gluon-gluon and quark-antiquark collisions as a function of the
    effective power-law dependence, $n_\text{eff}$, of the partonic
    luminosities on the system mass squared.}
  \label{tab:fxx-neff}
\end{table}

The normalisations of Eqs.~(\ref{eq:4}) and (\ref{eq:10}) have been
chosen such that the contents of the square brackets in each can be
directly compared.
Specifically, the relative sizes of the contents of the square
brackets reflect the relative sizes at LO of the
$d\sigma/d(\frac12 m_{\ttbar})^2$ and $d\sigma/d(\pTt)^2$ distributions.
Taking the working point
$\frac12 m_{\ttbar} \simeq \pTt \simeq 1\TeV$, where the $gg$ and
$q\bar q$ partonic luminosities are comparable for
$\sqrt{s} = 13\TeV$ (cf.\ Table~\ref{tab:channels}), and taking into account $n_\text{eff}\simeq 2$
from Fig.~\ref{fig:rap-structure-fixed-pTt} (right), we see that the
distribution of (half) the invariant mass is about six times larger
than that for the transverse momentum of a top quark.
Higher-order corrections will of course affect this result, but the
basic pattern of a strong (logarithmic) enhancement for the
$\frac12 m_\ttbar$ distribution relative to the $\pTt$ distribution
provides a helpful analytic explanation of the pattern seen in
Fig.~\ref{fig:parton-cross-sections-4-V}.

\section{Comments on the $m_\ttbar$  distribution beyond LO}
\label{sec:mttbar-beyond-LO}

\begin{figure}
  \centering
  \includegraphics[width=0.48\textwidth,page=1]{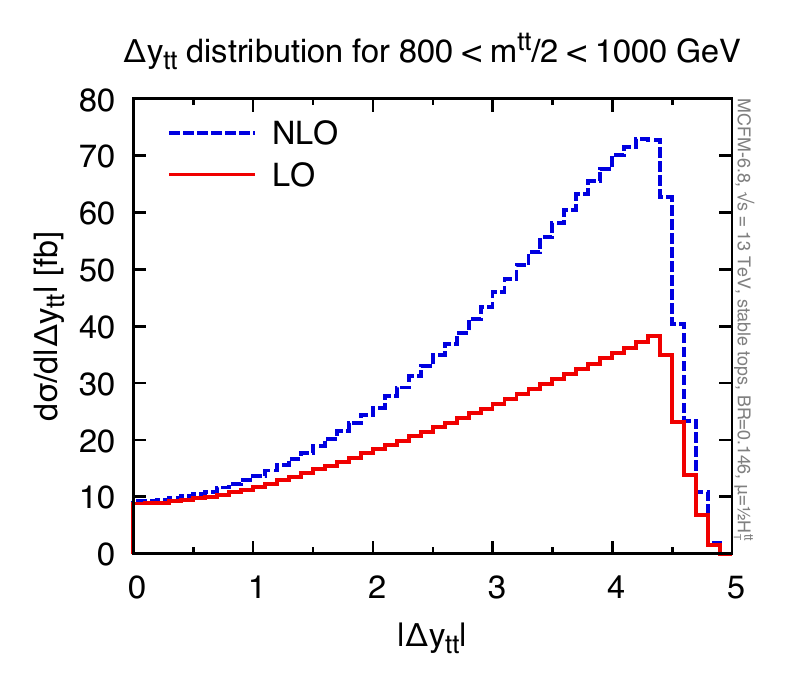}
  \includegraphics[width=0.48\textwidth,page=2]{plots/dytt-fixed-mtt-NLO.pdf}
  \caption{
    Left: NLO v. LO prediction for the $\dytt$ distribution in a bin
    of large $m^\ttbar/2$, as obtained with the MCFM program~\cite{Campbell:1999ah},
    including the semi-muonic branching fraction.
    The renormalisation and factorisation scales are set to $\mu =
    H_T^{\ttbar}/2$.
    Right: corresponding K-factor, i.e.\ the ratio of the NLO to the
    LO result. 
  }
  \label{fig:dytt-fixed-mtt-NLO}
\end{figure}

Fig.~\ref{fig:dytt-fixed-mtt-NLO} compares LO and NLO fixed-order
calculations (obtained using MCFM~\cite{Campbell:1999ah}) with a renormalisation
and factorisation scale choice of $H_T^{\ttbar}/2$.
This scale choice already absorbs the running coupling and PDF
effects associated with the varying top transverse momentum across the
range of $\dytt$ values (cf.\ the left and right-hand plots of
Fig.~\ref{fig:rap-structure}).
Those effects correspond to an $\as^3 \ln^2 m_\ttbar/\mtop$
contribution to the total cross section, i.e.\ a single-logarithmic
enhancement. 
However Fig.~\ref{fig:dytt-fixed-mtt-NLO} shows further strong
dependence of the NLO/LO $K$-factor as a function of $\dytt$, which
one may take evidence of further sources of logarithmic enhancement. 

A full discussion of the different potential sources of logarithmic
enhancement is significantly beyond the scope of this paper.
However, we believe that it is still informative to outline the
different classes of term that can contribute.
Loss of top momentum through fragmentation can contribute logarithms
at small $\dytt$ (i.e.\ large $p_T^t$), and is traditionally accounted
for in the FONLL formalism~\cite{Cacciari:1998it,Cacciari:2018qlp}.
At large $\dytt$, for $t$-channel top-quark exchange there are single
logarithmic $t$-channel-fermion analogues of BFKL
enhancement~\cite{Sen:1982xv,Fadin:1976nw,Fadin:1977jr,Bogdan:2006af},
associated with integrals over the rapidity of emitted gluons between
the two final-state top quarks.
Integrating over $\dytt$, we also expect double logarithmic
$\as^n \ln^{2n} m_\ttbar/\mtop$ enhancements, which relate to
double-logarithmic non-singlet structure functions at
small-$x$~\cite{Kirschner:1983di,Ermolaev:1995fx}, whose formalism can
be used~\cite{Blumlein:1995jp} to predict the 
$\as^n \ln^{2n-2}x$ terms in the non-singlet $P_\text{NS}^+(x)$
splitting functions~\cite{Curci:1980uw,Moch:2004pa,Moch:2017uml}. 
\logbook{a7f7deb}{see ../small-x-NS/evln-eqn.nb for checks}
To understand their origin in the context of $\ttbar$ production, one
may examine the NLO case: consider an ISR $g\to\ttbar$ splitting
followed by a harder $tg \to gt$ scattering (which proceeds mainly
through $t$-channel top exchange).
We are interested in a situation where, in the ISR splitting, the
anti-top and top have transverse momenta equal to some value $p_{t1}$.
Take the anti-top to be emitted into the final-state, and the top to be the
particle that initiates the $tg \to gt$ hard scattering.
The anti-top and top carry longitudinal momentum fractions $1-z$ and
$z$ respectively, with $z\ll 1$.
The $tg \to gt$ scattering itself involves large
$s,|u| \gg |t|\gg p_{t1}^2$.
The $zP_{g\to\ttbar}(z)$ splitting function goes as $\as z$, and the
reduced cross section $d\sigma/d\ln m_{tg}$ for the $tg \to gt$
scattering goes as $\frac{\as^2}{m_{tg}^2} \ln \frac{m_{tg}}{p_{t1}}$,
where $m_{tg}^2 = z m_\ttbar^2$.
The $z$ factor from the splitting function compensates the $1/z$ from
the $1/m_{tg}^2 = 1/(z m_{\ttbar}^2)$ factor in the reduced cross section, resulting in a
logarithmic integral over $z$.
There is also a logarithmic integral over $p_{t1}$.
Thus, after all integrations, we have a total cross section that goes
as $\as^3 \ln^3 m_{\ttbar}/\mtop$, i.e.\ a double-logarithmic
enhancement relative to the LO cross section.

Starting from NNLO, additional contributions arise that involve
$t$-channel gluon exchange, associated with four-top production.
At sufficiently large $m_\ttbar$ these will dominate the $m_\ttbar$
distribution because $d\sigma/dm_{\ttbar}^2$ scales as $\as^4/(\mtop^2 m_\ttbar^2)$
instead of the $\as^2/m_\ttbar^4$ seen in
Eq.~(\ref{eq:sigma-mtt-LO}).\footnote{To quantify this, care is needed
  in specifying a definition of $m_\ttbar$ when there are four tops
  from which to choose.}
There are also electroweak contributions from $q\bar q \to \ttbar$
through a $t$-channel $W$ exchange, which is dominated by 
$q\bar q = b\bar b$ incoming flavours. 
These scale as $\alpha_\text{EW}^2/(\mtop^2 m_\ttbar^2)$, with a
further $\as^2 \ln^2 \mtop/m_b$ suppression factor coming from the
requirements of two $b$-PDFs rather than gluon PDFs.
These $t$-channel vector-boson exchange contributions, whether they
involve gluons or electroweak bosons, will additionally be enhanced
by QCD~\cite{Kuraev:1977fs,Balitsky:1978ic} and
EW~\cite{Bartels:2006kr} BFKL-type logarithms.
We have verified the size of $4$-top and EW processes using
\texttt{Alpgen}~\cite{Mangano:2002ea} v.~2.14 and
\texttt{MadGraph5$\_$aMC@NLO}~\cite{Alwall:2014hca} v.~2.8.2
respectively.
At the LHC, even at large $m_\ttbar$ values of up to $4\TeV$, both
processes bring small corrections relative to the QCD LO $\ttbar$
cross section, at most at the few-percent level.
At a $100\TeV$ collider, which is beyond the scope of this
  article, considering $m_\ttbar \ge 20 \TeV$, the
  $1/\mtop^2 m_\ttbar^2$ scaling results in the $4$-top and EW
  processes becoming comparable to normal QCD $\ttbar$ production.
\logbook{30ffa45}{ For plots of $p_t$ and $\Delta y$ distributions see
  ../alpgen/4Qwork-gavin/results/4t.pdf
\begin{verbatim}
## Alpgen 4top results with mu=mtop=172.5
## and PDF set=MSTW2008nlo
## 
## [U] next to a result means it looked unstable
## in the vegas iterations
##
## Raw results (from separate different runs) and 
## input card files are to be found in the results/
## directory
rts     maxmtt_min  maxpt_max   sigma_gg(pb)     sigma_qq
13000    2000         300       (8.1 +- 0.1)e-7   (2.7+-0.1)e-6
130000   20000        300        failed to give an answer
13000    2000          --       (4.84 +- 0.03)e-4   (3.8+-0.1)e-5
130000   20000         --       (3.33 +- 0.06)e-3   (1.1+-0.1)e-5
1300000  200000        --       (4.06 +- 0.09)e-3   (5.7+-0.7)e-7[U]
13000    4000          --       (4.33 +- 0.08)e-6   (2.8+-0.1)e-7
100000   20000         --       (6.5  +- 0.1 )e-4   (3.0+-0.2)e-6

## Pythia8.303 ttbar results (with 4C tune)
## all decay channels
rts     mtt_min  sigma(pb)
13000    2000     0.21
130000   20000    2.6e-3
1300000  200000   2.9e-5
13000    4000     6.6e-4
100000   20000    6.2e-4

## MadGraph 2.8.2
## b b~ > t t~ QCD=0 (dominated by W exchange)
## scale choices need cross-checking
## NNPDF23_lo_as_0130_qed
rts     mtt_min  sigma(pb)
13000    2000     3.7e-3
130000   20000    3.7e-3
1300000  200000   3.8e-3
13000    4000     3.5e-5
100000   20000    8.2e-4

## cross checking s s~ > t t~
rts     mtt_min  sigma(pb)
13000    2000     2.3e-6
130000   20000    9e-8 +- 1e-8
1300000  200000   runs into trouble [did not try hard]
\end{verbatim}
}

We have also explicitly verified the stability of the $\dytt$
distribution from NLO to NNLO using the MiNNLO event sample of
Ref.~\cite{Mazzitelli:2020jio} across a range of $m^{\ttbar}/2$
values.\footnote{We are grateful to the authors for providing us with
  the event sample.}
\logbook{ec4d44a}{../ttjet-analysis/2020-12-results/hvq-v-minnlo-deltay-LO-NLO-MINNLO.pdf
  and ../ttjet-analysis/2020-12-results/hvq-v-minnlo-deltay.pdf}
Relative to a fixed-order (MCFM) NLO calculation with a scale choice
of $\mu = m^{\ttbar}/2$ that is similar to that of the MiNNLO sample,
we see a further substantial correction from the NNLO terms at large
$\dytt$.
It would be interesting to further investigate this with a scale
choice such as $\HT/2$ or $m_{T}^{J,\text{avg}}$ that tracks the
transverse momenta of the tops across the full $\dytt$ range.

We encourage further investigation of all the issues discussed in this
Appendix.

\bibliographystyle{JHEP}
\bibliography{tops}

\end{document}